\documentstyle[11pt,aaspp4]{article}

\def\kms{km s$^{-1}$}

\lefthead{Ferrarese et al.}
\righthead{Calibration of Population II Secondary Distance Indicators.}

\begin{document}

\title{The $HST$ Key Project on the Extragalactic Distance Scale XXVI.}

\title{The Calibration of Population II Secondary Distance Indicators
and the Value of the Hubble Constant}

\centerline{Accepted for publication in the {\it Astrophysical Journal}}

\author{Laura Ferrarese\altaffilmark{1,2},
Jeremy R. Mould\altaffilmark{3}, 
Robert C. Kennicutt, Jr.\altaffilmark{4}, 
John Huchra\altaffilmark{5}, 
Holland C. Ford\altaffilmark{6}, 
Wendy L. Freedman\altaffilmark{7}, 
Peter B. Stetson\altaffilmark{8}, 
Barry F. Madore\altaffilmark{9}, 
Shoko Sakai\altaffilmark{10},
Brad K. Gibson\altaffilmark{11}, 
John A. Graham\altaffilmark{12}, 
Shaun M. Hughes\altaffilmark{13}, 
Garth D. Illingworth\altaffilmark{14},  
Daniel D. Kelson\altaffilmark{12},
Lucas Macri\altaffilmark{5}, 
Kim Sebo\altaffilmark{3}, 
\& N.A. Silbermann\altaffilmark{9}  
}

\altaffiltext{1}{Hubble Fellow}
\altaffiltext{2}{California Institute of Technology, Pasadena CA 91125, USA}
\altaffiltext{3}{Research School of Astronomy \& Astrophysics, Institute of
Advanced Studies, ANU, ACT 2611, Australia}
\altaffiltext{4}{Steward Observatory, The University of Arizona, Tucson AZ 85721,
USA}
\altaffiltext{5}{Harvard Smithsonian Center for Astrophysics, Cambridge MA 02138
USA}
\altaffiltext{6}{Johns Hopkins University and Space Telescope
Science Institute, Baltimore MD 21218, USA}
\altaffiltext{7}{Carnegie Observatories, Pasadena CA 91101, USA}
\altaffiltext{8}{Dominion Astrophysical Observatory, Victoria, British Columbia
V8X 4M6, Canada}
\altaffiltext{9}{NASA/IPAC Extragalactic Database and California Institute of
Technology, Pasadena CA 91125, USA}
\altaffiltext{10}{Kitt Peak National Observatory, NOAO, Tucson AZ 85726, USA}
\altaffiltext{11}{CASA, University of Colorado, Boulder, CO, USA}
\altaffiltext{12}{Department of Terrestrial Magnetism, Carnegie Institution of
Washington, Washington DC 20015, USA}
\altaffiltext{13}{Royal Greenwich Observatory, Cambridge CB3 OHA, UK}
\altaffiltext{14}{Lick Observatory, University of California, Santa Cruz CA 95064
USA}

\begin{abstract}

A Cepheid-based calibration is derived for four distance
indicators that utilize stars in the old stellar populations: the tip
of the red giant branch (TRGB), the planetary nebula luminosity
function (PNLF), the globular cluster luminosity function (GCLF) and
the surface brightness fluctuation method (SBF). The calibration is
largely based on the Cepheid distances to 18 spiral galaxies within
$cz =1500$ km s$^{-1}$ obtained as part of the $HST$ Key Project on the
Extragalactic Distance Scale, but relies also on Cepheid distances
from separate $HST$ and ground-based efforts. The
newly derived calibration of the SBF method is applied to obtain
distances to four Abell clusters in the velocity range between 3800
and 5000 \kms, observed by Lauer et al. (1998) using the
$HST$/WFPC2. Combined with cluster velocities corrected for a
cosmological flow model, these distances imply  a value of the Hubble
constant of 

$$ H_0 = 69 \pm 4 {\rm (random)} \pm 6 {\rm
(systematic)~km~s^{-1}~Mpc^{-1}.}$$

This result assumes that the  Cepheid PL relation is independent of
the metallicity of the variable stars; adopting  a metallicity
correction as in Kennicutt et al. (1998), would produce a (5 $\pm$
3)\% decrease in $H_0$.  Finally, the newly derived calibration allows
us to investigate systematics in the Cepheid, PNLF, SBF, GCLF and
TRGB distance scales.  

\end{abstract}


\section{Introduction}

Peculiar and infall velocities in the Local Supercluster are
comparable to the Hubble flow.  Consequently, $H_0$ cannot be
determined by simply measuring the distances to a random assortment of
nearby galaxies at velocities of a few hundred \kms.  In view of this,
the $H_0$ Key Project  (Kennicutt, Freedman \& Mould 1995, Freedman et
al. 1998) was designed with three primary goals in mind: 1) to use a
high quality primary standard candle, Cepheid variable stars, in
nearby galaxies to calibrate reliable secondary distance indicators
that are luminous enough to reach galaxies with heliocentric
velocities up to  $\sim 10000$ \kms, well beyond any substantial
Hubble flow deviations. 2) To provide a  check on  potential
systematic errors both in  the Cepheid distance  scale and the
secondary methods.   The Cepheid distance database from the $HST$ Key
Project can  be used   both to  calibrate well-studied secondary
methods, as  well   as  to  test  the  accuracy of other   suggested
indicators,  and ultimately to calibrate them  if they are found to be
reliable. 3)  To make direct  Cepheid measurements of  distances  to
three spiral galaxies in each of the Virgo and Fornax clusters.

Our primary goal of using the Cepheid period-luminosity (PL)
relation to measure accurate distances to 18 carefully selected
galaxies in the $3-20$ Mpc range has been accomplished.  We now turn
our attention to the calibration of secondary distance indicators for
extension into the smooth Hubble flow. The need for a multiplicity of
calibrators goes beyond square-root-n considerations: much of the
current uncertainty in the distance scale stems from the likelihood of
systematic errors that are not fully understood, together with
uncertainties in the dispersion of secondary distance indicators,
which affects their use through Malmquist bias. Other papers in this
series will calibrate the Tully-Fisher relation (Sakai et al. 1999),
the $D_n$-$\sigma$ relation and the fundamental plane (Kelson et al.
1999), and the Type Ia supernovae (Gibson et al. 1999).   In this
paper we use Cepheid distances to calibrate four additional distance
indicators:  the luminosity of the tip of the red giant branch (TRGB),
the planetary nebula luminosity function (PNLF), surface brightness
fluctuations (SBF), and the globular cluster luminosity function
(GCLF).  Because all of these secondary candles  utilize stars in old
stellar populations, we imprecisely refer to them as Population II
distance indicators, without implying that the stars are necessarily
metal poor.  The calibrations in subsequent sections will illustrate
the relative strengths and weaknesses of the four Pop II distance
indicators, and show where more work must be done on particular
indicators.

Of the distance indicators discussed in this paper, at present only
the SBF method has the potential of reaching the unperturbed Hubble
flow.  Thanks to the  $HST$ Near Infrared Camera and Multi Object
Spectrometer (NICMOS) galaxies as far as $cz = 10000$ \kms~are within sight
(Jensen et al. 1999). SBF measurements currently point to values of
$H_0$ around 80 \kms~Mpc$^{-1}$.  In particular, velocities and
distances to the four Abell clusters in the 3800$-$5000 \kms~velocity
range observed with the $HST$/WFPC2 by Lauer et al. (1998), while
leading directly to $H_0=82 \pm 8$ \kms~Mpc$^{-1}$, also allowed the
calibration of the far field brightest cluster galaxies Hubble diagram
of Lauer and Postman (1992), giving  $H_0 = 89 \pm 10$
\kms~Mpc$^{-1}$. In a very recent preprint, Tonry et al. (1999) combined
their latest calibration of the ground-based $I$-SBF method with a
sophisticated model for large scale velocity flows to derive $H_0 = 77
\pm 8$ \kms~Mpc$^{-1}$.  Finally, Thomsen et al. (1997) derive $H_0=71
\pm 11$ \kms~Mpc$^{-1}$ from the SBF distance to NGC 4881 in the Coma
cluster.

The usefulness of TRGB, PNLF and GCLF  methods resides mainly in
providing cross checks as to the reliability of each indicator. In
addition, all of the distance indicators presented in this paper have
the indisputable merit of being applicable to both spiral and
elliptical galaxies. As such they serve as a much needed link between
the Cepheid distance scale and early type galaxies. This link cannot
be provided by the Tully-Fisher relation, nor the $D_n-\sigma$ and
the fundamental plane, which are exclusive to spirals and ellipticals
respectively.

TRGB distance estimates are routinely performed in the Local Group,
even if the use of $HST$ has brought the Leo I group, and even Virgo
into sight (Sakai et al. 1997, Ferguson et al. 1998, Harris et
al. 1998).  GCLF measurements beyond the Virgo and Fornax clusters are
technically extremely challenging,  as testified by the $HST$
observations of the Coma galaxy IC 4051 (Baum et al. 1995, 1997). The
wide range of $H_0$ values derived from GCLF measurements in the Virgo
and Fornax clusters, ranging from the mid 50s (Sandage \& Tammann
1996) to the low 80s \kms~Mpc$^{-1}$ (Grillmair et al. 1998), can be
traced back to uncertainties in the peculiar and infall velocities for
these clusters, and to the still rather uncertain calibration of the
method itself.  Finally, PNLF measurements become prohibitive for
galaxies beyond $\sim 40$ Mpc even using 8-m class telescopes
(Jacoby 1998). Distances to Virgo and Fornax suggest values of $H_0$
in the upper 70s to mid 80s \kms~Mpc$^{-1}$ (Jacoby, Ciardullo \& Ford
1990, McMillan, Ciardullo \& Jacoby 1993, Ford et al. 1996)

As a first step in calibrating the Pop II distance indicators
presented in this paper, we compiled a database of all available
Cepheid, TRGB, PNLF, GCLF and SBF measurements. The results are
published separately in Ferrarese et al. (1999, hereafter F99). Here
we will use the distance database not only to provide a Cepheid
calibration for TRGB, PNLF, GCLF and SBF, but also to test each
method, and the Cepheid distance scale  itself, for biases.  This
paper is organized as follows.  A brief overview of the Cepheid
calibration adopted and the steps taken to prepare the database for
the derivation of the magnitude zero points, are discussed in \S
2. The distance indicators are then considered, from the least to
the most far reaching ones: TRGB, PNLF, GCLF and SBF are
calibrated in \S 3  though \S 6, and the results discussed in \S
7. There we will compare the different methods and  discuss  the
magnitude of the metallicity dependence of the Cepheid PL relation,
and its effects on the calibrations derived in the previous sections.
The newly calibrated SBF distances to clusters beyond $cz = 3800$ \kms~are
discussed in \S 8, and a value for the Hubble constant is derived.  A
summary of conclusions can be found in \S 9.

\section{Cepheid Distances and Preparation of the Database}

A compilation of all Cepheid distances obtained as part of the $H_0$
Key Project, as well as independent ground-based and $HST$ efforts
is given in F99. In that paper, all distances are placed on a
homogeneous system, with consistent calibration and fitting of the PL
relation, treatment of the extinction corrections, and estimate of the
errors.  Here we will only remind the reader that the PL relation is
calibrated  on a sample of 32 LMC Cepheids with  $BVI$ photoelectric
photometry and periods in the range $1.6 < P < 63$ days (Madore \&
Freedman 1991), and assumes  a true distance modulus and average
line-of-sight reddening  to the LMC of $18.50 \pm 0.13$ and $E(B-V) =
0.10$ mag respectively,  a ratio of total to selective absorption
$R_V=A(V)/E(B-V)=3.3$, and a reddening law following Cardelli, Clayton
and Mathis (1989).  The derived distances do not appreciably depend
on variations in $R_V$, or on the adopted $E(B-V)$ and reddening law
for the LMC (Ferrarese et al. 1996, F99)

Those galaxies having distances that had to be modified to adhere to
the criteria listed above  (SMC, NGC 3109, NGC 1365, NGC1425, NGC
1326A, NGC 4535) and galaxies with distances that will not be 
considered for calibrating secondary distance indicators (NGC 2403, GR8,
NGC 2366, and NGC 4571) are discussed in detail in F99.

Because no consensus has yet been reached on the issue of a
dependence of the Cepheid PL relation on the metallicity
of the variable stars (e.g. see Kennicutt et al. 1998), we prefer not
to include metallicity effects explicitly in the Cepheid distances
listed in the F99 database, but we will discuss the impact of a
metallicity dependence on the distances and the calibration of
secondary distance indicators in \S 7.2.

In the F99 database,  distance moduli are given only for the Cepheid
measurements. For the other indicators, magnitudes (the magnitude of
the tip of the RGB, the cutoff magnitude of the PNLF, the turnover
magnitude of the GCLF and the fluctuation magnitude for SBF), rather
than distances, are given. To provide a consistent dataset, these
magnitudes are not corrected for reddening. Unlike the Cepheids
(e.g. Freedman et al. 1994)  the secondary distance indicators can
only be corrected for the effects of foreground extinction, as no
estimate of the internal reddening is available.  This is not as big a
handicap as it might appear as regions suitable for SBF measurements,
PNe, TRGB stars and globular clusters are all found in areas
relatively free of internal dust absorption (Tonry \& Schneider 1988,
Feldmeier, Ciardullo \& Jacoby 1997, Whitmore 1996).  Estimates of the
foreground reddening $E(B-V)$ are available from the HI maps of
Burstein and Heiles (1984), or from the 100 $\mu$m maps reprocessed
from IRAS/ISSA and COBE/DIRBE data by Schlegel et al. (1998). The two
methods show a systematic difference in the values of $E(B-V)$ of
about 0.02 mag (the DIRBE/IRAS maps giving higher reddenings), the
cause of which is not yet understood. In this paper, the DIRBE/IRAS
maps have been adopted: they have higher angular resolution, provide a
more direct measure of the dust column density, and are more accurate
than the HI maps, especially in regions of moderate and high
reddenings. For comparison, magnitude zero points derived using
Burstein \& Heiles reddenings will also be presented in the
Appendix. As a word of caution, we need to point out that recent
findings claim that reddenings from DIRBE/IRAS maps might be overestimated by up
to 50\% in  regions of high extinction [$E(B-V) > 0.15$ mag] (Arce \&
Goodman 1999). Because  the Galactic reddening for all of our galaxies
is smaller than $E(B-V) \sim 0.1$ mag, these findings do not directly
concern our analysis. The reddening law used to convert $E(B-V)$
values to extinction in different passbands is from Cardelli, Clayton
and Mathis (1989). The value of $R_V$ for the diffuse interstellar
medium is $R_V=3.1$, which is generally derived from early type stars
(Schultz \& Wiemer 1975, Whittet \& van Breda 1980, Rieke \& Lebofsky
1985). To account for the convolution of the broad band filters and
the redder spectral energy distribution of the Pop II objects
considered in this paper (globular cluster, elliptical galaxies) a
value of $R_V=3.3$ has been adopted for all distance indicators with
the exception of PNLF. PNe are observed in a narrow band filter,
therefore $R_V=3.1$ is the appropriate value to use\footnotemark.
Note that all of the distance indicators considered in this paper are
little affected by internal extinction (the contribution of which is
neglected), therefore the low internal $R_V$ claimed for elliptical
galaxies ($\sim 2.1-3.3$, Goudfrooij et al.  1994) is not an issue. A
treatment of errors associated with the  extinction corrections is
given in Appendix A.

\footnotetext{For $R_V=3.3$,
the extinction curves of Cardelli et al.  give the following ratios
for the absorption: $A(B):A(5007):A(V):A(I):A(F814W):A(K_{s}):A(K')$ =
1.288:1.120:1:0.600:0.596:0.120:0.125. }

\section{Calibration of the Tip of the Red Giant Branch Method}

Reviews of the tip of the red giant branch (TRGB) method as a distance
indicator are given by Madore, Freedman \& Sakai (1996) and Sakai (1999).
Briefly, the TRGB marks the He ignition in the degenerate core of low
mass stars. The core mass is constant for stellar ages larger than  a
few Gyrs, thereby producing a constant luminosity for the Helium
flash. The absolute $I$ magnitude of the TRGB proves to be fairly
invariant over a wide range of metallicity (Da Costa \& Armandroff
1990, Lee, Freedman \& Madore 1993, Salaris \& Cassisi 1998), making
the TRGB a powerful standard candle.

Unlike the PNLF, GCLF and SBF methods, TRGB is not currently
calibrated using Cepheids.  Two calibration philosophies are being
pursued, leading to a 0.15 mag discrepancy in the absolute $I$
magnitude of the tip (Lee, Freedman \& Madore 1993, Salaris \& Cassisi
1998).  In the range of metallicities  $-2.35 <$ [M/H] $< -0.28$ dex
[corresponding to $1.2 < (V-I)_{-3.5} < 2.2$ mag, where $(V-I)_{-3.5}$
is the color measured at an absolute $I$ magnitude of $-3.5$], the
Salaris and Cassisi calibration produces a  quadratic dependence of
$M_I^{TRGB}$ on metallicity, with  $-4.25 \le M_I^{TRGB} \le -4.15$
mag. In contrast, Lee et al. find that $M_I^{TRGB}$ varies  between
$-3.9$ and $-4.1$ mag for $-2.17 <$ [Fe/H] $< -0.71$ [$1.2 <
(V-I)_{-3.5} < 1.8$ mag]. The discrepancy is due entirely to different
estimates of the TRGB bolometric luminosity, which Salaris \& Cassisi
adopt from theoretical stellar models (Salaris \& Cassisi 1996), while
Lee et al. adopt from the empirical observations of Galactic globular
clusters with RR-Lyrae distances (Da Costa \& Armandroff 1990,
Frogel, Persson \& Cohen 1983). The latter calibration elevates the
TRGB to the rank of primary distance indicator, similar to the
Cepheids.

In this paper, we will treat the TRGB as a secondary distance
indicator, and provide a calibration through the Cepheids. This will
allow us to perform a consistency check between
the primary TRGB calibration and the Cepheid distance
scale. Furthermore, the relative insensitivity of the TRGB to
metallicity, and the lack of correlation between the halo and Cepheid
metallicity,  provides an independent way to test the metallicity
dependence of the Cepheid PL relation (Lee et al. 1993, Kennicutt et
al. 1998), this will be discussed in \S 7.2.  For present purposes, we
do not apply any metallicity correction  to the TRGB magnitudes. Over
the range of metallicities spanned by the TRGB calibrators [1.1$ <
(V-I)_{-3.5} <$ 1.6 mag], the Lee et al.  and Salaris \& Cassisi
estimates disagree not only on the magnitude of such correction (which
are however  small, within 0.1 mag), but also on their sign.

From F99, all galaxies with both Cepheid distances and an estimate of
the TRGB $I$ magnitude belong to the Local Group: LMC, Sextans A and
B, IC 10, IC 1613, M31, M33, NGC 3109 and NGC 6822. For these galaxies,
a Cepheid $-$ TRGB  comparison is shown in Figure 1  (assuming no
metallicity dependence of the Cepheid PL relation or the TRGB magnitudes), 
and  leads to a magnitude zero point 

$$M_I^{TRGB}=-4.06\pm0.07~ {\rm( random)}~\pm 0.13~ {\rm (systematic)}~ {\rm mag.} \eqno{(1)}$$ 

We must stress that the large  errors in the  $(V-I)$ colors prevent
us from quantifying a color (metallicity) dependence of $M_I^{TRGB}$,
and therefore equation (1)  applies correctly only at the sample mean
$(V-I)\sim 1.42$ mag.  The fit leading to equation (1) is done
accounting for errors in both Cepheid distances and TRGB magnitudes;
the random error is the 1$\sigma$ error in the fit, while the
systematic error is the combination, in quadrature, of the uncertainty
in the LMC distance (0.13 mag) and the uncertainty in the zero point
of the LMC PL relation (0.02 mag, see \S8). Adding to the above
calibrators the (almost) direct comparison between the Cepheid
distance to M81 and the $M_I^{TRGB}$ for the M81 dwarfs BK5N and F8DI,
produces the minor adjustment $M_I^{TRGB}=-4.05\pm0.06$ mag (random
error only).  A calibration of $M_I^{TRGB}$ based on group distances
is not as robust as the calibration given in equation (1), because
TRGB observations outside the Local Group and the M81 group are
available to only one galaxy in each of the Leo I group, the  M87
subcluster in the Virgo cluster (see F99 for a discussion of the
spatial subdivision of the Virgo cluster into separate subclusters)
and the NGC 5128 group, and therefore cluster depth effects are likely
to affect the group distances. Using group distances to the four
groups mentioned above, not weighted by their spatial extent, would
lead to $M_I^{TRGB}=-4.00\pm0.05$ mag, in good agreement with the
result given in equation (1).  The magnitude zero points discussed in
this section and in \S 4$-$6 are listed in Table 1. In that table, the
reduced $\chi^2$ of the fit can be found in parentheses next to each
zero point; the underline identifies the magnitude zero points which
we consider most trustworthy for each indicator.

Figure 2  shows a comparison between the secondary calibration
presented in equation (1) and the primary calibration of the TRGB.
The top panel shows the color dependence of the absolute $I$ magnitude
of the TRGB predicted by Lee et al. (dashed line) and Salaris and
Cassisi (the solid and dotted lines differ exclusively for the adopted
bolometric correction, the dotted line is based on the stellar models
of Kurucz 1979, the solid line on the Yale isochrones). 
The lower panel shows the Cepheid $-$ TRGB distance moduli
residuals when the calibration given in  equation (1) is applied
(solid dots and error bars). It is quite obvious that the large
uncertainty in distance moduli (and in colors, which are not plotted
in the figure) prevent testing the correlation between the residuals
and the colors.  The Lee et al.  calibration would lead to the
residuals shown by the open circles, with a mean shown by the dashed
line, while the Salaris \& Cassisi calibration (simply for
illustration we chose the Kurucz models) produces the residuals shown
by the stars, with a mean given by the dotted line.  The Cepheid
calibration is virtually indistinguishable from the Lee et al. (1993)
primary calibration, and agrees with the Salaris \& Cassisi (1998)
calibration at the 1.2$\sigma$ level (we assume a 0.1 mag uncertainty
for the primary calibrations).  Therefore, Salaris \& Cassisi's claim
that the 0.1$-$0.2 mag level disagreement between their TRGB
calibration and the Cepheid distance scale is statistically
significant  seems premature at this point, in view of the large
uncertainties. Furthermore, the perfect agreement between the Lee et
al.  calibration (which bypasses the Cepheids completely, relying only
on RR-Lyrae distances to Galactic globular clusters) and the Cepheid
calibration (based on an assumed distance to the LMC), is a strong
reassurance as to the  reliability of our adopted LMC distance and error.

\section{Calibration of the Planetary Nebula Luminosity Function Method}

Reviews of the Planetary Nebula luminosity function (PNLF) and its
application as a distance indicator are given by Jacoby et al. (1992),
Jacoby (1996), Mendez (1998) and Jacoby, Ciardullo \& Feldmeier
(1998).  PNe owe their role of standard candles to the universality of
their luminosity function.  Because PNe are high excitation objects,
the LF is measured at the [OIII]$\lambda$5007 emission line, and it is
approximated by a semiempirical formula which combines theoretical
models with the sharp cutoff at the bright end (characterized by a
`cutoff' magnitude $m^*$) observed for the PNLF in the bulge of M31
(Ciardullo et al. 1989). Effects of metallicity, population age,
Hubble type and absolute luminosity of the parent galaxy are claimed
to be negligible (Jacoby \& Ciardullo 1999; Ciardullo \& Jacoby 1992;
Jacoby 1996; Dopita et al. 1992; Stanghellini 1995; Jacoby, Ciardullo,
\& Harris 1996; McMillan, Ciardullo \& Jacoby 1993; Jacoby, Walker \&
Ciardullo 1990).  Potential sources of biases are contamination by HII
regions in spiral galaxies, by background emission line galaxies at $z
= 3.1$ (Jacoby, Ciardullo \& Harris 1996, Ciardullo et al. 1998), and by
intracluster PNe (Ciardullo et al. 1998). HII regions should not
present a serious threat since they can be discriminated from PNe
based on their H$\alpha$/[OIII] ratio; however intracluster PNe can
lead to an underestimate of the PNLF cutoff magnitude (and therefore
the PNLF distance) by up to 0.2 mag in rich clusters (Mendez 1998).
PNLF measurements are performed routinely in both spiral and
elliptical galaxies reaching as far as the Virgo and Fornax cluster
(Jacoby, Ciardullo \& Ford 1990, McMillan, Ciardullo \& Jacoby 1993,
Ciardullo et al. 1998), but unfortunately exposure times become
prohibitive for galaxies beyond $\sim 40$ Mpc even using 8-m class
telescopes (Jacoby et al. 1998).

The absolute magnitude corresponding to $m^*$ is currently  calibrated
solely by the Cepheid distance to M31 from Madore \& Freedman (1991),
giving $M^* = -4.54$  mag (at $\lambda = 5007$ \AA, e.g. Ciardullo et
al. 1998).   However, the PNLF has now been measured in six galaxies
with Cepheid distances (F99). These are the LMC, M31, M81, M101, NGC
300 and NGC 3368 in the Leo I group.  A comparison of Cepheid distance
moduli and PNLF cutoff magnitudes $m^*$ for these galaxies is shown in
Figure 3, and produces an absolute cutoff magnitude 

$$M^*=-4.58 \pm 0.07 {\rm ~(random)} \pm 0.13  {\rm ~(systematic)}{\rm ~mag}, \eqno{(2)}$$ 

\noindent assuming no metallicity dependence of the Cepheid PL
relation.  As for all other secondary distance indicators presented in
this paper, the fit is performed accounting for errors in both
coordinates. Because the calibration of PNLF is based on both
ground-based and $HST$ Cepheid galaxies (unlike TRGB, which is based
exclusively on ground-based galaxies, and GCLF and SBF which are
based exclusively on $HST$ galaxies), a 0.09 mag error has been added
in quadrature to the distance moduli for the $HST$ galaxies prior to
performing the fit. This term accounts for the (in this case) random
error associated with the $HST$ photometry, and is not included
explicitly in the error on the distance moduli quoted in F99. The
calibration in equation (2), which we consider the most robust  to
date, reproduces the M31-based calibration of Ciardullo \& Jacoby
within 0.04 mag, in spite of the different Galactic reddenings adopted.
There are five groups/clusters with both PNLF and Cepheid mean
distances. These are the NGC 1023 group, the Leo I group, the M87 and
NGC 4472 subclusters, and  the Fornax cluster; the  NGC 4649
subcluster in the Virgo cluster has been excluded as it deviates by
more than 3$\sigma$ from the best fit. The indirect comparison of PNLF
magnitudes and Cepheid distances to these groups (excluding the Leo I groups for which there is a direct comparison) can be added to the
direct calibrators. To provide meaningful weighting,  a measure of the
group/cluster depth (based on the projected extent of the group in the
sky as listed in Table 2), is added to the uncertainty in the group
mean Cepheid distances. This leads to a zero point of  $M^*=-4.61 \pm
0.06$ mag, consistent with the above adopted calibration.

\section{Calibration of the Globular Cluster Luminosity Function Method}

Reviews of the GCLF method can be found in Harris (1991), Harris
(1996) and Whitmore (1996). The use of globular cluster systems for
distance determinations relies on the empirical result that their luminosity
function is remarkably well approximated by a Gaussian (e.g. Abraham
\& van den Bergh 1995) with the turnover $M_T$ acting as a standard
candle. The calibration of $M_T$ is still rather controversial (Secker
1992, Sandage \& Tammann 1995, Whitmore et al. 1996, Della Valle et
al. 1998); furthermore, reliable estimates of the GCLF turnover
magnitudes require deep photometry that sample the GCLF well over the
turnover. The latter requirement is due to the fact that the
dispersion of the GCLF is not universal, but rather shows a
large scatter both within late type spirals (e.g. in the $V$ band
$\sigma=1.10$ mag for M31 while $\sigma=1.42$ mag for the Milky Way,
Secker 1992, see also the result obtained by Bothun et al. 1992 on NGC
7814), and early type galaxies (e.g. $\sigma=1.67$ mag for the E0 NGC
5481, Madejsky \& Rabolli 1995, while $\sigma=1.09$ mag for the E1 NGC
4494, Forbes 1996).

The existence of second parameters affecting $M_T$ has been widely
discussed in the literature. There are claims that $M_T$ depends on
the luminosity of the parent galaxy, brightening  by about 0.3 $V$
magnitudes going from dwarf systems to giant ellipticals (Whitmore
1996); on the environment, being fainter for galaxies in rich clusters
(Blakeslee \& Tonry 1996);  and on Hubble type, being $\sim 0.15~ V$
mag brighter in spirals than ellipticals (Whitmore 1996).  This last
observation has been attributed by Ashman, Conti \& Zepf (1995) to
metallicity variations.  There is evidence that the luminosity
function of the metal poor and metal rich clusters do not peak at the
same magnitude in the sense of the blue clusters being brighter (Elson
\& Santiago 1996, Kundu et al. 1999, Kundu and Whitmore
1998). Finally, contamination by a spatially extended GC component
associated with tidal debris, and not necessarily linked to any
particular galaxy (much as is the case for intracluster PNe, see \S
4), should be taken into account (West et al. 1995).

Traditionally, GCLF measurements have been performed preferentially in
the Johnson $B$ or $V$ bands. The lack of color
information, and the (not  well defined) dependence of globular
cluster  color on the luminosity of the host galaxy calls for an
independent calibration of the GCLF in these two bands.  The most
up-to-date calibration of the $V$ band GCLF is by Whitmore  et
al. (1996). Based on the weighted means of the turnover magnitudes for
six Virgo ellipticals and the Cepheid distances to NGC 4321, NGC 4496A
and NGC 4536, the absolute $V$ magnitude of the GCLF turnover is found
to be $M_V^T=-7.21 \pm 0.26$ mag. This calibration is in agreement
with the one derived by  Secker (1992) based on RR Lyrae distances for
the Milky Way globular clusters, $M_V^T = -7.29 \pm 0.13$ mag (but see
also Sandage \& Tammann 1995); however, Secker (1992) also notes that
a calibration based solely on the distance to M31 (to date the only
galaxy with both a GCLF and Cepheid distance) would lead to the
discrepant value $M_V^T = -7.69 \pm 0.15$ mag (Secker's result was
modified here to adopt the M31 Cepheid distance from Madore and Freedman 1991.)

The Whitmore et al. calibration can now be improved using the F99
database, which includes a more homogeneous and reliable set of GCLF
measurements (many GCLF turnover magnitudes used by Whitmore are
derived from data that do not properly sample the turnover, and a few
are based on $B$, rather than $V-$band measurements) and a larger
sample of Cepheid distances.

An indirect calibration of the $V-$band GCLF can be derived using
group distances to  the  M87 subcluster, the  NGC 4472
subcluster, and the Fornax cluster. The GCLF turnover magnitude is
defined by M87 and NGC 4478 for the  M87 subcluster, and by NGC
4472 for the NGC 4472 subcluster. The turnover magnitudes are
within 0.03 mag, as expected if the two subclusters are at
approximatively the same distance as supported by the PNLF and SBF
observations presented in this paper (see also Gavazzi et al. 1999).
Since M87 is at the bottom of the potential well of the cluster
(B\"ohringer et al. 1994), M87 and NGC 4472 should give a good
estimate of the cluster mean distance. The mean turnover magnitude of
the Virgo cluster is 23.68$\pm$0.06 mag in $V$, 0.2 mag brighter than
estimated by Whitmore (1996). The Cepheid distance modulus to Virgo,
defined by NGC 4321, NGC 4496A, NGC 4536, NGC 4548 and NGC 4535 is
31.03 $\pm$ 0.03 mag, or 0.07 mag fainter than the value adopted by
Whitmore. The resulting $V$ turnover magnitude for Virgo is
$-7.35\pm$0.07 mag.  Since Whitmore's calibration, Cepheid and GCLF
distances have become available for the Fornax cluster. The GCLF mean
turnover magnitude to Fornax is defined by NGC 1344, NGC 1380, NGC
1399 and NGC 1404, while the mean Cepheid distance is defined by NGC
1326A, NGC 1425 and NGC 1365. This gives a turnover magnitude of
$-7.85\pm$0.07, {\it a full 0.5 magnitudes brighter than that found
for the Virgo cluster}.  Of the four GCLF galaxies in Fornax, NGC
1344, NGC 1399 and NGC 1380 are found within 0.06 mag of each other,
while NGC 1404 is 0.28$\pm$0.14 mag in the background,  but still
consistent with it belonging to the cluster, for which we estimate a
depth of $\sim 0.3$ mag from the projected extent of the cluster in
the sky.  A formal weighted mean of the Virgo and Fornax absolute
magnitudes of the  turnover,  for no metallicity correction of the
Cepheid PL relation, gives (Figure 4):

$$M_V^T = -7.60 \pm 0.25 {\rm ~(random)} \pm 0.16 {\rm~ (systematic)} {\rm ~mag}. \eqno{(3)}$$

The random error quoted in the equation above reflects the actual
scatter seen in the data, rather than the formal uncertainty in the
fit. The observed scatter is incompatible with
the published internal errors in the GCLF measurements, and the
expected uncertainty due to cluster depth effects, pointing to the
existence of second parameters.  The systematic error is a combination
of the uncertainty in the PL calibration, the error in the distance to
the LMC and the error affecting the $HST$ photometry (see \S 8). The
latter are of a systematic rather than random nature, because all
Cepheid  galaxies used for the calibration of GCLF were  observed with
the same $HST$ instrumental configuration.  The magnitude zero point
in equation (3) is consistent  with the result  obtained using a
comparison of Cepheid and GCLF distances for M31, $M_V^T = -7.70 \pm
0.19$  mag, however the discrepancy between the $M_V^T$ values for
Virgo and Fornax is worrisome, and is evident in the large reduced
$\chi^2$ of the fit ($\sim$12, from Table 1). It is difficult to
imagine a scenario in which such a large discrepancy could be due to a
different spatial distribution of the Cepheid and GCLF galaxies within
the two clusters.  The sense of the discrepancy is such that the
Cepheid galaxies would be foreground to the Virgo cluster, and
background to the Fornax cluster, assuming the centers of the clusters
to be defined by the GCLF galaxies; however, both PNLF and SBF sample
the central ellipticals in Virgo and Fornax, and give a consistent
Virgo-Fornax relative distance of $\sim 0.35$ mag, with Fornax being
further, while GCLF places Fornax and Virgo at practically the same
distance.  Based on the current evidence, we conclude that at present,
the GCLF method carries an unexplained uncertainty that may be quite
large. The limited amount of data makes it difficult to establish
whether this is due to the fact that the GCLF itself is not a standard
candle, or to problems in the published GCLF measurements, for example
arising from incorrect estimates of the incompleteness of the data or
contamination by background galaxies.

The situation for the $B-$band GCLF is more precarious still, as the
sample of available measurements is very limited.  The only direct
calibration available in the literature for the $B-$band GCLF was
provided by Harris et al. (1991) based on 75 halo clusters with RR
Lyrae distances, giving $M_B^T = -6.84 \pm 0.17$ mag.  There are four
groups for which a mean Cepheid distance and GCLF $B$ turnover
magnitude can be defined: the NGC 4472 and NGC 4639 subclusters in
Virgo, the Fornax cluster and the Leo I group.  A formal fit gives
(Figure 4):

$$M_B^T = -7.02 \pm 0.5 {\rm ~(random)} \pm 0.16 {\rm ~(systematic)} {\rm ~mag}. \eqno{(4)}$$

As for the case of the calibration of $M_V^T$, the random error
reflects the actual scatter seen in the data. As was the case for the
$V-$band GCLF,  the points show an uncomfortably large dispersion:
$M_B^T$ is $-6.33\pm$0.23 mag, $-6.98\pm$0.10 mag, and $-7.48\pm$0.26
mag, for Virgo, Fornax and Leo I respectively.    Excluding the NGC
4649 subcluster, for which we find in  \S 4 and \S 6  a large Cepheid
distance  compared to both PNLF and SBF measurements, leads to $M_B^T
= -6.95 \pm 0.5$ mag.  

\section{Calibration of the Surface Brightness Fluctuation Method}

The surface brightness fluctuation (SBF) method was first introduced
by Tonry \& Schneider (1988); reviews can be found in Jacoby et
al. (1992), Tonry et al. (1997) and Blakeslee, Ajhar \& Tonry
(1999). The method can be applied to both elliptical galaxies and the
dust-free bulges of  early-type spirals. The fluctuation
magnitude, $\overline{m}$, measures the amplitude of the  luminosity
fluctuations arising from the counting statistics of the stars
contributing to the flux in each pixel of a high S/N image.
$\overline{m}$ is heavily dominated by RGB stars, and therefore its
absolute calibration $\overline{M}$ is strongly wavelength and color
(metallicity) dependent. 

To date, the largest SBF survey is  the ground-based effort by Tonry
et al. (1999) in the Kron-Cousins $I-$band, which comprises over 300
galaxies within $cz = 4000$ \kms. With $HST$/WFPC2 and the F814W
filter, data extending to the Coma cluster  have been obtained
(Thomsen et al. 1997). In the future, the emphasis of SBF is expected
to shift to the near-infrared, where SBF magnitudes are brighter and
dust contamination is less threatening (e.g. Blakeslee, Ajhar \& Tonry
1999).  A few galaxies have been observed in the $K'$ ($\lambda_c =
2.10 \mu$m) and $K_{s}$ ($\lambda_c = 2.16 \mu$m) bands (Jensen et
al. 1998, Pahre \& Mould 1994), and  using $HST$/NICMOS (Jensen et
al. 1999). Because of the small size of these IR surveys (the
$HST$/NICMOS data have not been published at the time of writing this
paper), we limit the present discussion to the $I$ and F814W SBF, and
touch briefly on the $K'$ and $K_s$ surveys in Appendix B.  Five dwarf
elliptical galaxies in the Sculptor group have been observed in the
$R-$band  by Jerjen, Freeman \& Binggeli (1998). Because of the lack
of overlap between this sample and any other Cepheid or  SBF sample,
the $R-$band SBF must be calibrated using stellar population synthesis
models, and will not be discussed in  this paper.

\subsection{$I-$band SBF}

A Cepheid-based calibration of the $I$-SBF method is given by Tonry
et al. (1999) as  $\overline{M}_I = (-1.74 \pm 0.07) + (4.5 \pm
0.25)[(V-I)_0 - 1.15]~ {\rm mag}$, where $(V-I)_0$ is the galaxy
unreddened color determined in the same region where the fluctuations
are measured. The zero point is derived by Tonry et
al. using the median for six galaxies with SBF measurements and
Cepheid distances (from F99).   The slope of the $\overline{M}_I$  vs
$(V-I)_0$ relation  was determined by  Tonry et al. (1997) by
simultaneously fitting all of the galaxies in groups and clusters for
a single value of the slope and a different zero point for each
group. The slope thus derived is in prefect agreement with the
predictions of stellar population models (Worthey 1993, 1994),
which also predict a zero point of $-1.81$ mag, in good agreement with
Tonry et al. (1999). 

For the purpose of this paper, we do not attempt to recalibrate the
color dependence of the SBF magnitude,  but rather adopt the 4.5 slope
from Tonry et al. (1997).   The galaxies with both Cepheid distances
and $I$-SBF magnitudes are M31, M81, NGC 3368 in Leo I, NGC 4725 in
the Coma II cloud, NGC 4548 in the Virgo cluster, and NGC 7331. A
comparison for these galaxies is shown in Figure 5, where the
fluctuation magnitudes are corrected for extinction (see \S 2), and
for the color dependence. A weighted least-squares fit gives:

$$\overline{M}_I = [-1.79\pm0.09 {\rm (~random)} \pm 0.16 {\rm
(~systematic)}] + (4.5 \pm 0.25)[(V-I)_0 - 1.15] {\rm
~mag}. \eqno{(5)}$$  

The difference between the zero point above and the one derived by
Tonry et al. (1999) is due to two factors. Tonry et al. use the
extinction law from Schlegel et al. (1998) rather than adopting the
same law used in deriving the Cepheids distances (from Cardelli
Clayton \& Mathis 1989). Second, Tonry et al. (1999) use a median of
the zero points given by all calibrators rather than a weighted
mean. This effectively rejects NGC 7331 which not only produces the
brightest zero point (Figure 5), but also carries a large weight in a
our weighted mean, having smaller error bars than the other
galaxies. Choosing between the zero point presented in this paper and
the one given by Tonry et al. (1999) is largely a matter of personal
preference, the two agreeing well within one sigma. However, the
calibration in equation (5) is to be preferred when comparing the
secondary distance indicators presented in this paper, since it shares
the same methodology used for TRGB, PNLF and GCLF.

To avoid uncertainties introduced by cluster depth effects, we prefer
the direct calibration in equation (5) to a group calibration (such as
the one adopted by Tonry et al. 1997), which would lead to a 0.1 mag
($1\sigma$) dimming of $\overline{M}_I$ (based on the NGC 1023 group,
the Fornax cluster, the Leo I group, the Virgo M87 and NGC 4472
subclusters, the NGC 5128 group, and the NGC 7331 group). On the
downside, a  direct calibration relies on the assumption of an
identical stellar population in the bulges of the spiral calibrators
and the ellipticals which are the main targets of the method. 
Additionally,  as testified by the large error bars in the fluctuation
magnitudes of bulges, SBF measurements in spirals are made difficult
by the presence of dust and  irregularities in the luminosity
profile. The latter point is responsible for the larger dispersion
shown in Figure 6 by the spirals compared to the ellipticals.  In the
lower part of the figure we follow  Tonry et al. (1997) and fit all
galaxies belonging to the  five groups listed for five different zero
points and a common slope. The solid points represent spiral galaxies
in each group: given the admittedly limited amount of data, the
spirals do not seem to occupy a different locus in the
[$(V-I)_0,\overline{M}_I$] plane compared to the ellipticals, which
speaks in favor of a common $\overline{M}_I$ for bulges and
early-type galaxies.  To illustrate the play of cluster depth
effects, the upper part of Figure 6 shows the same galaxies but  this
time shifted vertically according to the mean Cepheid distance of the
group to which they belong. Compared to the lower part of the figure,
the points have now moved upwards (to fainter magnitudes) relative to
the six spiral calibrators represented by the filled pentagons, due to
galaxy type segregation in the clusters. 

Based on the SBF and Cepheid distances, the elliptical and spiral
populations in the Leo I group and the Fornax cluster have similar
back-to-front extent: $\sim 1$ Mpc for the Leo I group and $\sim
5$ Mpc for the Fornax cluster. While the mean distance for both
populations in Fornax is $\sim 21$ Mpc, the ellipticals in Leo I seem
to be $\sim 1$ Mpc in the background of the spirals, which give a mean
distance for the group of 10.5 Mpc. The situation is more complicated
for the Virgo cluster. The ellipticals in the M87 and NGC 4472
subclusters show a fairly homogeneous distribution between 15 and 20
Mpc,  with a mean close to 17 Mpc, while the Cepheid spirals show a
significantly smaller back-to-front spread of less than 1 Mpc at
a mean distance of 16 Mpc.  Both SBF and Cepheid distances to the NGC
4649 subcluster do not support physical association for the galaxies
in this region: SBF distances show a large back-to-front spread of
0.8 mag ($\sim 8$ Mpc), while the Cepheid distance puts NGC 4639 at
least 2 Mpc in the background of the region defined by the ellipticals.

\subsection{$HST/$F814W SBF}

Of the distance indicators presented in this paper, only F814W SBF has
targeted galaxies beyond the 2000 \kms~velocity range, which is of
interest in the determination of the Hubble constant.  A calibration
of F814W SBF is therefore of great importance.  The most recent
calibration for F814W SBF is presented by Ajhar et al. (1997). The
small sample size and the almost complete absence of overlap with
Cepheid galaxies make it necessary to derive both zero points and
slope of the color dependence of F814W SBF by comparison with the
$I$-SBF distance moduli for the galaxies observed in both photometric
bands. Using a sample of 16 galaxies with $I$-SBF distance moduli
(Tonry et al. 1997, 1999), Ajhar et al. derive: $ \overline{M}_{F814W} =
(-1.73 \pm 0.07) + (6.5 \pm 0.7)[(V-I)_0 - 1.15]~ {\rm mag}.$

The galaxies in common between the F814W and $I$-SBF samples are
listed in F99 and shown in Figure 7, where we have used equation (5)
in calibrating the $I$-SBF magnitudes. Using a two-parameter fit,
accounting for errors in both coordinates, gives $\overline{M}_{F814W}
= (-1.91\pm0.10) + (7.4 \pm 1.2)[(V-I)_0 - 1.15]$ mag, and is shown by
the dashed line in Figure 7. However, we prefer to adopt a different
procedure than the one followed by Ajhar et al. (1997): the
Kron-Cousin $I$ and the WFPC2/F814W filters are not very dissimilar,
and there is no reason to expect significant changes in the stellar
population as seen in the two passbands. In fact,  stellar population
models (Worthey 1993, 1994) predict a slope of 4.3 for
$\overline{m}_I$ in the color range $1.05 < (V-I)_0 < 1.35$
(Blakeslee, Ajhar \& Tonry 1999),  close to the slope observed by
Tonry et al. (1997). The same models applied to $\overline{M}_{F814W}$
predict a virtually identical slope of 4.25 (Blakeslee, private
communication). We conclude that the present amount of data is too
limited to allow an empirical determination of 
the slope of the color dependence, and we prefer to impose the $I-$band
slope of $4.5 \pm 0.30$ on $\overline{M}_{F814W}$ (note that the error
has been inflated compared to equation (5)  to account for the 0.05
mag difference between the theoretical calibration in the two
bands). This produces the fit shown by the solid line in Figure 7:

$$\overline{M}_{F814W} = [-1.70 \pm 0.10 {\rm ~(random)} \pm 0.16 {\rm
~(systematic)}] + (4.5 \pm 0.30)[(V-I)_0 - 1.15]~ {\rm mag} \eqno{(6)}$$

As a consistency check, we also derived a magnitude zero point for
F814W SBF using a comparison with Cepheid group distances for the Leo
I group, the NGC 1023 group, and the  M87 and NGC 4472 subclusters
(the discrepant  NGC 4649 subcluster has been excluded), after
color correcting the F814W fluctuation magnitude using the slope given
in equation (6). The resulting magnitude zero point, $-1.50 \pm 0.04$
mag, is  $2\sigma$ fainter than the one  given by equation (6),
confirming the importance of depth effects in deriving mean distances. 

To conclude we must stress that the calibration of F814W
SBF is likely to be significantly improved in the near future, when more F814W SBF measurements will be performed for galaxies in the $HST$ public
archive. In particular, it would be desirable to derive the slope of
the color dependence using a large sample of galaxies belonging to
different groups and clusters, as was possible for $I$-SBF (Tonry et
al. 1997).

\section{Discussion of Distance Scales}

\subsection{Comparison of Secondary 
Distance Indicators}

A summary of the magnitude zero points discussed in \S 3 through \S 6
is presented in Table 1, based on a direct (`Gal' in column 3), group
(`Group') or mixed (`Gal/Group') comparison with the Cepheid distance
moduli.  The number of calibrators for each method is shown in column
2.  Zero points are derived both assuming no metallicity dependence of
the Cepheid PL relation (column 4) and a dependence as in Kennicutt et
al. (1998) (column 5); the two sets of zero points will be compared in
\S 7.2.  Our final adopted zero points are underlined in the table.
Applications of these zero points to the database of F99 gives the
TRGB, PNLF, GCLF and SBF distances listed in Table 2. Distances
derived from near infrared SBF measurements are followed
by a colon  to stress the fact that they are of limited reliability
(see Appendix B).  In Table 2,  galaxies are grouped
according to their cluster association, defined in F99. Column 2 gives
the `membership index' as in F99: galaxies which are certain, very
likely, or just probable members of their assigned group or cluster
are designated as class 1, 2, and 3 respectively. The last column in
the table gives a weighted mean of all distances available to any
particular galaxy, with the exclusion of GCLF, $K_s$ and $K'$ SBF.
Heading each group of galaxies, is the cluster designation and the
weighted mean of the cluster distances based on PNLF, GCLF, $I$-SBF,
F814W-SBF and TRGB measurements to its members.  In the case of
$I$-SBF, more galaxies are averaged in the group distance than shown
in the table. The full sample of galaxies  can be found in Ajhar et
al. (1999).  Also noted in the table is the depth, in magnitudes, of
each cluster, estimated based on the spatial extent of the cluster in
the sky (see F99).

The only Cepheid calibrator in common to all secondary distance
indicators is M31; therefore a comparison between the TRGB, PNLF, GCLF
and $I$-SBF distance scales is not circular. Such a comparison is
shown in Figures 8 and 9, and the results of a weighted
least-squares fit to the data are summarized in  Table 3.

Given the fact that the calibration of the $V-$band GCLF is rather
uncertain, the only serious systematic effect emerging from Figures 8
and 9  is the discrepancy in zero points between the PNLF and SBF
distance scales, also reported by Mendez (1998). We have tested for
correlations between the distance moduli residuals and the galaxy
color, as used in the SBF calibration, absolute $B$ magnitude,
T-type, and Mg$_2$ index, and  found nothing of significance.  Note
that adopting a metallicity dependence of the Cepheid PL relation
leading to larger distances to metal rich galaxies, as in Kennicutt et
al. (1998), worsens the situation (see Table 1).  A comparison between
PNLF and Cepheid distance moduli (Figure 3 and Table 2) also shows a
systematic offset between the local PNLF calibrators (which only
extend as far as the Leo I group) and the Virgo and Fornax
clusters. In a recent review,  Jacoby, Ciardullo \& Feldmeier (1998)
found that PNLF distances to Virgo ellipticals are, on average, 1.3
Mpc closer than Cepheid distances to Virgo spirals. Our study confirms
Jacoby et al's result, and it also shows a similar, but more
pronounced trend for the Fornax cluster, the mean PNLF distance to
which is 3.6 Mpc closer than measured by the Cepheids. This discrepancy
is only partially reconciled if NGC 1425 is assumed to be in the 
background of the cluster: in this case the ellipticals with PNLF distances
would be 1.8 Mpc closer than the two Cepheid spirals, NGC 1326A
and NGC 1365.

There are several possibilities to explain the above observations.
One is depth effects combined with Hubble type segregation in the
Virgo and Fornax clusters.  Indeed, Jacoby et al's preferred
explanation is that elliptical galaxies in  the Virgo cluster are
foreground with respect to the spirals. We do not favor this
conclusion in view of the even larger PNLF$-$Cepheid discrepancy
observed in the more compact Fornax cluster. Furthermore, PNLF and SBF
distances {\it to the same early-type galaxies} in Virgo and Fornax
are systematically offset by 0.3 mag, with the SBF galaxies giving
mean cluster distances in better agreement with the Cepheids. A more
likely cause of the observed systematics, in our opinion,  is a bias
in one, or more, of the  PNLF, SBF or Cepheid distance scales. 

While we cannot exclude Cepheids and SBF as being in error, three
observations lead us to believe that PNLF is at least partially to
blame: 1) as mention above, Cepheid and SBF distances agree for both
the local sample and the Virgo and Fornax clusters, arguing that PNLF,
rather than SBF or the Cepheids, is more likely affected by a
systematic bias.  2) Recently, Mendez (1998)  pointed out that
contamination by intracluster PNe (stripped PNe that `float' in the
potential well of the cluster, but are not associated with any
particular galaxy) could lead to an underestimate of the  PNLF
distances by up to 0.2 mag, particularly in moderately rich
environments such as Virgo and Fornax. This effect acts in the same
sense and is of the right order of magnitude to explain the PNLF/SBF
and PNLF/Cepheid discrepancy.  Because the Virgo and Fornax clusters
in our sample also happen to be the richest ones, contamination of
intracluster PNe would be more severe for these clusters than for the
more nearby and smaller groups, as observed. 3) TRGB distance moduli 
agree well within 1$\sigma$ with the SBF distance moduli, while 
the agreement with PNLF is marginally worse (1.5$\sigma$).

We have considered a few more possibilities that could lead to the
observed systematic offsets, with no success. These include a
metallicity dependence affecting the Cepheid distances, a possible
corruption of the SBF measurements due to dust patches, stellar
aggregates or globular clusters,  and an inadequate corrections for
Galactic reddening affecting both the PNLF and SBF
distances. The metallicity dependence of the Cepheid PL relation has
already been exonerated.  The presence of dust or other contaminants
in the SBF galaxies would lead to an overestimate of the fluctuation
magnitudes, and thereby an underestimate of the distances, which is
contrary to the observations. Because the Galactic extinction is
practically the same within a particular cluster, if the amount of
reddening in Virgo and Fornax had been overestimated this would
artificially bring the PNLF distances closer, and the SBF distances
further (because of the color dependence of the SBF
calibration). However,  even if the adopted DIRBE/IRAS reddenings are
higher than the Burstein \& Heiles (1984) reddenings, adopting the
latter would bring things into better agreement by no more than a few
hundredths of a magnitude.

Finally, we tested for possible correlations between residuals of any
combination of distance indicators and T-type, absolute $B$
luminosity of the host galaxy, distance and environment (as given by
the group velocity dispersion from Tully 1988), and  found nothing of
significance. In particular, this applies to previous claims of a
dependence of the GCLF turnover magnitudes on environment and of the
PNLF cutoff magnitude on the B magnitude of the host galaxy (Figure
10), reported by Blakeslee \& Tonry (1996) and Bottinelli et
al. (1991) respectively. A Spearman rank-order correlation test
gives a significance level consistent with the null hypothesis of no
correlation within 1$\sigma$ in both cases.  Note that Blakeslee \&
Tonry's claim was mainly motivated by published GCLF turnover
measurements in Leo and Coma (Harris 1990, Baum et al. 1997). These
measurements are not used in this paper, because they are based on
GCLFs that do not adequately sample the turnover.

\subsection{Effects of a Possible Metallicity Dependence of the Cepheid PL Relation}

A review of the  metallicity dependence of the Cepheid PL relation can
be found in Kennicutt et al. (1998). To date, no agreement has 
been reached as to the magnitude of the change in the Cepheid distance
modulus $(m-M)_0$ as a function of Cepheid metallicity [O/H]: some
estimates which apply to Cepheid distances derived from $V$ and $I$
observations are $\delta(m-M)_0/\delta[O/H] < -0.1$ mag dex$^{-1}$
(Chiosi, Wood \& Capitanio 1993), $\sim -0.4$ mag dex$^{-1}$ (Sasselov
et al. 1997, Beaulieu et al. 1997, Kochanek 1997) and $\sim -0.24$ mag
dex$^{-1}$ (Kennicutt et al. 1998), in the sense of increasing the
distances of galaxies more metal rich than the LMC, which calibrates
the Cepheid PL relation. However, recently Bono et al. (1999) derived
$\delta(m-M)_0/\delta[O/H] \sim 0.75$ mag which acts in the {\it
opposite}  sense, i.e. decreasing the distance to galaxies more metal
rich than the LMC.   Because of the lack of consensus, in the
previous sections we preferred to derive magnitude zero points for the
secondary distance indicators applying no metallicity correction in
the Cepheid PL relation.  The magnitude of such correction can be
judged from Figures 1 and 3$-$5, and Table 1 where we list magnitude
zero points for both $\gamma = \delta(m-M)_0/\delta[O/H] = 0,  {\rm
~and~} \gamma = -0.24$ mag dex$^{-1}$.  While the effects of the
metallicity dependence on the Cepheid distance to any single galaxy
can be considerable,  the repercussion on the calibration of secondary
distance indicators is generally attenuated by having calibrators
spanning a wide range in metallicities.  Generally, galaxies with
Cepheid distances outside the Local Group are large spirals more metal
rich than the LMC. For example, the Cepheid distance to Fornax and
Virgo increase by 0.05 and 0.15 mag respectively when a metallicity
dependence as in Kennicutt et al. (1998) is assumed. Therefore, the
calibration of PNLF, GCLF and SBF, which is heavily weighted by
galaxies outside the Local Group, will show a corresponding moderate
dependence on the assumed metallicity dependence of the Cepheid PL
relation, of the order of 0.05 to 0.10 mag for $\gamma= 0.24$ mag
dex$^{-1}$.  In contrast, with the exception of M31 and  M33, all of
the Local Group galaxies with Cepheid distances are irregular systems
slightly more metal poor than the LMC. The calibration of TRGB, which
is based mainly on Local Group galaxies, shows a 0.05 change in zero
point with metallicity, but in the opposite sense to that shown by
PNLF, GCLF and SBF.

The data provided here can be used to check for systematic residuals
versus Cepheid metallicity, as given by the [O/H] indexes.  Such a
test performed using TRGB distances is particularly effective
(Kennicutt et al. 1998).  The correlation between TRGB $-$ Cepheids
distance moduli residuals and [O/H] indexes is shown in Figure 11.
The best weighted least-squares fit is consistent with a correlation
at the 1.5$\sigma$ level, and gives a slope of $-0.18\pm0.15$ mag
dex$^{-1}$.  Bearing in mind that this test cannot be considered
definitive, in view of the large error bars, the limited amount of
data, and the lack of data points at high metallicities, this result
is consistent with a metallicity dependence as in  Kennicutt et
al. (1998) and is smaller than predicted by  Sasselov et al. (1997)
and Beaulieu et al.  (1997).  While a similar test performed
using $I$-SBF shows no significant correlation, the PNLF$-$Cepheids
residuals correlate with the Cepheids [O/H] indexes at the 2$\sigma$
level,  but in opposite sense than found above. This is unlikely due
to a  metallicity dependence of  the PNLF distances (Jacoby \&
Ciardullo 1999), and we rather favor the conclusion of a spurious
result, pinned by only one galaxy, NGC 3368, at the high metallicity end.

\section{Addressing the Hubble Constant: Application of SBF}

Of the distance indicators presented in this paper, F814W-SBF
reaches far enough into the unperturbed Hubble flow to warrant its
application in estimating the Hubble constant.  Lauer et al. (1998)
produced F814W-SBF measurements to the central galaxies in the Abell
clusters A262, A3560\footnotemark, A3565, and A3742, with
heliocentric velocities between 3800 and 4900 \kms.  An F814W
fluctuation magnitude for NGC 4881 in the Coma cluster is published by
Thomsen et al. (1997); however, the galaxy was not observed in the $V$
band, and therefore the color dependence of the fluctuation magnitude
cannot be calculated. Thomsen et al. estimate the amount of
metallicity correction to $\overline{M}_{F814W}$ based on a rather
uncertain relation between $\overline{m}_{F814W}$ and the Mg$_2$
index, calibrated using only one galaxy, NGC 3379. Finally, Pahre et
al. (1999) present F814W-SBF measurements to NGC 4373 in the NGC
4373 group in the Hydra-Centaurus supercluster, at a heliocentric
velocity of $\sim 3400$ \kms.  

\footnotetext{The galaxy observed by Lauer et al., attributed to Abell
3560, in reality belongs to a group at $cz = 3806$ \kms~in the foreground of
the cluster.  The actual Abell 3560 cluster is at a heliocentric
velocity of $\sim 15000$ \kms~(Willmer et al. 1999).}

Reddenings, colors and fluctuation magnitudes for these galaxies are
summarized in F99. For NGC 4881, we assumed the same color as for NGC
3379, since the two galaxies share the same Mg$_2$ index (Thomsen et
al. 1997) and added a large error bar to incorporate the range of
colors observed for early type galaxies in the SBF database: $(V-I)_0
= 1.2 \pm 0.1$.  Distance moduli are calculated using the zero point
given in equation (6), and the results are presented in Table 4. 

The $\sim 0.15$ mag difference between our distance modulus for NGC
4373 and Pahre et al.  is due mainly to differences in the Galactic
extinction $E(B-V)$, our estimate being 0.03 mag smaller, and only
marginally to the different SBF calibration.  The distance moduli
derived for the four Abell clusters  are on average 0.3 magnitudes
(1.5$\sigma$) larger than those derived by Lauer et al. (1998).  Part
of this difference is due to the calibration, and part to the fact
that our adopted Schlegel et al. $E(B-V)$ reddenings are on average
0.03 magnitudes larger  than the Burstein et al. values used by Lauer
et al. Finally, the distance modulus to NGC 4881 in the Coma cluster
is hardly significant because of the large uncertainty associated with
the $(V-I)$ color; within the errors this distance is consistent with
the value derived by Thomsen et al.  (1997).

The last difficulty to overcome in our quest for $H_0$ is the
determination of the clusters' `cosmic' velocities. We adopt
velocities corrected for the local flow field as described in  Mould
et al. (1999).  Briefly, the heliocentric velocities (column 6 of
Table 4) are corrected first to the centroid of the Local Group using
the prescription of Yahil, Tammann \& Sandage (1977, listed in column
7 of Table 4; using other prescriptions, e.g. the IAU correction
listed in column 8, does not produce a substantial difference), and
then for infall towards three attractors, the contribution of which is
assumed to be additive: Virgo, the Great Attractor (GA), and the
Shapley Concentration. The flow-corrected velocities thus obtained
are listed in column 10 of Table 4.  For comparison we also list in
column 6 of the same table the heliocentric systemic velocity of the
cluster (from the CfA redshift Survey, Chen et al. 2000, see also
http://cfa-www.harvard.edu/\verb+~+huchra/clusters) and   in column 9
the velocities in the reference frame defined by the cosmic microwave
background (CMB), using the relation of Giovanelli et al.  (1998) and
a CMB dipole velocity as in Kogut et al. (1993).  In three cases the
differences between $v_{flow}$ and $v_{CMB}$ are larger than 10\% of
$v_{CMB}$. These are the clusters closest to the GA.  These
corrections make physical sense: the galaxies are moving towards the
GA, and, the question is, how fast? The model is not capable of making
accurate predictions close to the GA: linear infall is assumed, but
here we are well inside the non-linear regime (Schechter 1980).  Data
exist  on peculiar velocities for one of the three clusters, Abell
3560. Mould et al.  (1991) find a peculiar velocity of 970 $\pm$ 420
km/sec.  The model predicts 552 \kms. For these three clusters we
regard the flow model's predicted peculiar velocities as uncertain by
$\pm$ ($v_{CMB} - v_{flow})$ while for the remaining clusters we
assume an error of 300 \kms~following other papers in this series
(Sakai et al. 1999, Gibson et al. 1999, Kelson et al. 1999). 

The distance moduli and the flow-corrected velocities produce the
values of $H_0$  listed in column 11 of Table 4 and plotted in Figure
12. A final value of $H_0$ is derived by fitting the Hubble diagram
$v_{flow} = v_0 + H_0*d$, with $d$ equal to the distance to each
individual cluster, by constraining the intercept $v_0 = 0$, and
accounting for errors in both distances and velocities. This gives
$H_0=70 \pm 4$ \kms~Mpc$^{-1}$, where the error reflects only the
1$\sigma$ uncertainty in the fit. Notice that a bivariate fit would
not affect the value of $H_0$ and would give $v_0=75\pm1071$ km/s, which
justifies our imposed constraint of a zero intercept.  The value of
$H_0=70 \pm 4$ \kms~Mpc$^{-1}$ is heavily weighted by the four Abell
clusters; the Coma cluster and NGC 4373 carry  little weight because
of the  large error bars in their distances and velocities
respectively.  Note that the four Abell clusters yield $H_0$ that is
1$\sigma$ lower than that derived from NGC 4373, possibly suggesting
that the NGC 4373 group is in the foreground of the Hydra-Centaurus
supercluster, onto which it projects.

Table 5  gives a summary of the contributions to the final uncertainty
in the derived value of $H_0$. Errors in the Cepheid distance scale
have been reviewed extensively in all papers of this series; the
current presentation differs only in the separation of systematic
(identified by the letter $S$ in Table 5) and random ($R$)
contributions to the final error in the Cepheid distance to any
individual galaxy.   The errors in the
SBF distance scale include the  uncertainties in the photometric zero
point and the extinction correction, which are described in Appendix
A. Because the SBF fluctuation magnitudes are strongly dependent on
color (equation 6), the  error in the color-corrected F814W-SBF
magnitudes includes not only the random errors in the fluctuation
magnitudes (term 2.R$_1$), but also a term depending on the slope and
error of the color dependence (4.5 $\pm$ 0.3).    The color dependence
of the SBF magnitudes produces typical errors of 4.5 times the error
in the $(V-I)_0$ color (term 2.R$_2$), and 0.3
times [$(V-I)_0 - 1.15$], added in quadrature. All of the above errors
are random.  The only source of systematic error is the zero point of
the SBF calibration (term 2.S$_1$), $\pm$ 0.10 mag (\S 6.2). Note that
random errors in the Cepheid PL calibration (R$_{PL}$ in Table 5) are
already accounted for in computing the error on the SBF zero point,
and therefore need not be considered again.  The systematic error on
the Cepheid PL relation, however, has not yet been  accounted for, and
needs to be added in quadrature to term 2.S$_1$ to obtain the final
systematic error on the F814W SBF distance scale of 0.19 magnitudes
(S$_{SBF}$).

The error on the flow-corrected velocities (term 3.R$_1$) has been
discussed already. Errors on $H_0$ are given by the formulae listed in
the notes to part 3 of Table 5, for the case in which errors on the
velocities and distances ($d$ in Mpc) are identical for the $N$
galaxies used to  derive $H_0$; the generalization to the case in
which the errors differ from object to object (as is our case) is
straightforward.

In conclusion:

$$ H_0 = 69 \pm 4 {\rm (random)} \pm 6 {\rm (systematic)~km~s^{-1}~Mpc^{-1}}. \eqno{(7)}$$ 

\noindent as determined using only the four Abell clusters.  While we
stress that a larger sample of SBF measurements is required with
better sky coverage in order to reduce the propagation of the velocity
flow model  dependence into our estimate of $H_0$, we notice that
using  CMB, rather than flow velocities, does not sensibly affect our
result, producing $H_0 = 70 \pm 4$ km s$^{-1}$ Mpc$^{-1}$ (random
error only) as the weighted mean of the values obtained from the four
Abell clusters.  However, corrections for large scale flows are the
largest cause of the difference between the value of $H_0$ above and
the one derived by Tonry et al. (1999), $H_0 = 77 \pm 4 \pm 7$ km
s$^{-1}$ Mpc$^{-1}$.  In particular, the Tonry et al. (1999) and Mould
et al. (1999) flow models predict corrections of opposite sign for the
velocities  of two of the four Abell clusters considered in this
paper, Abell 3560 and Abell 3565, due to a different placement of the
Great Attractor.  This alone produces a 5.5\% increase in $H_0$ when
the Tonry et al. (1999) model is adopted. The remaining 5\% increase
is due equally to the adoption of a different zero point (see \S 6.1)
and to the largest sample of clusters beyond 3000 \kms~ available
to Tonry et al. (1999). In passing, we will also point out that the
difference between the value of $H_0$ derived in this paper and the
one presented by Lauer et al. (1998, from the same Abell cluster used
in our analysis) can be entirely ascribed to differences in the
Galactic reddenings and the adopted SBF calibration.

Finally, the effects of a metallicity dependence of both the Cepheid
PL relation, and the SBF magnitudes, need to be discussed. Assuming a
metallicity dependence of the Cepheid PL relation as in Kennicutt et
al. (1998) leads to a  $\sim (5 \pm 3)\%$ decrease in $H_0$ (open
circles in Figure 12).  In \S 6.2, we derived a calibration of F814W
SBF which differs from the original calibration of Ajhar et al. (1997)
both in the zero point and in the slope of the color dependence. We
also warned the reader that the latter is not  well defined given the
current set of calibrators. Fortunately,  $H_0$ is not very sensitive
to the slope of the color term in the calibration of F814 SBF, as long
as this is accompanied by the appropriate change in zero point.  For
example, adopting Ajhar et al's slope would produce only a $\sim$1\%
adjustment in $H_0$.

\section{Summary}
	
We have derived a Cepheid calibration for four secondary distance
indicators, applicable to the old stellar population in elliptical
galaxies and the bulges and halos of spirals: the tip of the red giant
branch method, the globular cluster luminosity function method, the
planetary nebula luminosity function method and finally the surface
brightness fluctuation method. GCLF and SBF data have been collected
in more than one photometric band, namely $V$ and $B$ for the GCLF,
and Kron-Cousins $I$, $K_s$, $K'$, and F814W for SBF, each of which
was calibrated independently.  TRGB, PNLF and $I$-SBF have been
measured in a sufficient number of galaxies with Cepheid distances
that their calibration can proceed through direct comparison for
individual galaxies.  However, the calibration of GCLF must
proceed through a comparison of  group, rather than individual
distances, while the calibration of F814W SBF is better done against
galaxies with $I$-SBF distances, calibrated in turn against the
Cepheids.

Our derived zero points, summarized in Table 1, are improved compared
to previously published values. This is not due only to the large
dataset of Cepheid distances employed in this paper, but also, and
perhaps mostly, to the homogeneity of the dataset used, both for the
Cepheids and for the secondary distance indicators.  While the
Cepheids, TRGB and SBF distance scales show good agreement throughout
the range of distances considered in this paper,  distances derived
using the PNLF method in the Virgo and Fornax clusters are
systematically smaller (by $\sim$0.35 mag) than given by either the
SBF method or the Cepheids. A conclusion as to the significance of the
discrepancy   between PNLF and Cepheid distances is  premature at this
point, since it would require a knowledge of the spatial  distribution
of early and late type galaxies in the notoriously complex Virgo
cluster. However, the difference between SBF and PNLF distances
measured {\it to the same galaxies} is real and needs to be solved
with future work. Contamination of intracluster PNe, which leads to
artificially overestimate the PNLF cutoff magnitudes especially in
richer clusters such as Virgo and Fornax, acts in the correct sense,
but is probably not sufficient, to correct the observed discrepancy.

We also found a large scatter in the distances derived using the GCLF
method, pointing to the existence of a second parameter affecting the GCLF turnover magnitude. At present, we estimate
the GCLF turnover magnitude to be uncertain by 0.25 mag and 0.5 mag in
the $V-$ and $B-$bands respectively. 

We studied the correlation of residuals of TRGB and Cepheid distances
against [O/H] indices as a test of the metallicity dependence of the
Cepheid PL relation. Our results are consistent with the findings of
Kennicutt et al. (1998) using Cepheid distances in two fields of
M101, and both are consistent with no metallicity dependence of the
Cepheid PL relation.  For PNLF and SBF, the effects of a metallicity
dependence of the Cepheid PL relation as in Kennicutt et al. (1998)
acts in the sense of brightening the magnitude zero points by about
0.10  magnitudes, due to the fact that most of the calibrating
galaxies are metal rich.  

We have applied our newly derived calibration to calculate F814W-SBF
distances to four Abell clusters with group velocities in the 4000
\kms~range, using the data by  Lauer et al. (1998). By adopting a velocity
flow model from Mould et al. (1999) to correct the cluster velocities
for large scale flows, we obtain a value for the Hubble constant of
$H_0 = 69 \pm 4 {\rm (random)} \pm 6 {\rm (systematic)}$
\kms~Mpc$^{-1}$. We find that this result is  insensitive
(at the 1\% level) to the slope of the metallicity dependence of the
F814W SBF fluctuation magnitude, however, a metallicity dependence of
the Cepheid PL relation as in Kennicutt et al. (1998) will lower the
Hubble constant by (5 $\pm$ 3)\%.

\acknowledgments

We wish to thank John Tonry, John Blakeslee, Ed Ajhar and Alan
Dressler for kindly giving us access to the SBF database prior to
publication.  Many thanks also to John Blakeslee, John Tonry, George
Jacoby, Pat C\^ot\'e and Pat Durrell for helpful discussions.  LF
acknowledges support by NASA through Hubble Fellowship grant
HF-01081.01-96A awarded by the Space Telescope Science Institute,
which is operated by the Association of Universities for Research in
Astronomy, Inc., for NASA under contract NAS 5-26555.  The work
presented in this paper is based on observations  with the NASA Hubble
Space  Telescope, obtained by  the   Space Telescope Science
Institute.  Support for this work was also provided by NASA  through
grant GO-2227-87A from STScI.  This research has made use of the
NASA/IPAC Extragalactic Database (NED), version 2.5 (May 27,
1998). NED is operated by the Jet Propulsion Laboratory,  California
Institute of Technology, under contract with the National  Aeronautics
and Space Administration.

\appendix

\section{Uncertainty in the Extinction Corrections}

Uncertainties in the extinction corrections contribute to the final
error in the dereddened magnitudes through: 1) the uncertain definition
of the extinction curve itself $k(\lambda - V) = E(\lambda
-V)/E(B-V)$; 2) errors in the assumed value of $R_V$; and 3) errors in
the adopted value of the foreground $E(B-V)$. In terms of the three
quantities above, $A_{\lambda}= (K(\lambda - V) + R_V) \times E(B-V)$;
therefore, the error on the extinction corrected magnitude is:

$$\sigma^2_{m_{\lambda}} = \sigma^2_{E(B-V)}\times (A(\lambda)/E(B-V))^2 +
E(B-V)^2 \times \sigma^2_{k(\lambda - V)} 
+ E(B-V)^2 \times \sigma^2_{R_V}  \eqno{(8)}$$

\noindent Similarly, $E(\lambda - V) = k(\lambda - V) \times E(B-V)$;
therefore, errors in the colors are given by

$$\sigma^2_{m_{(\lambda-V)}} = \sigma^2_{E(B-V)}\times k(\lambda-V)^2 +
E(B-V)^2 \times \sigma^2_{k(\lambda - V)}.  \eqno{(9)}$$

For wavelengths longwards of 4000 \AA, $\sigma_k(\lambda - V) \sim
0$ (Fitzpatrick 1999); therefore, the second term in the equations
above drops out.   The uncertainty in $E(B-V)$ is given as  $\pm 0.015$
mag for the Burstein and Heiles (1984) reddenings, and  16\% for the
DIRBE/IRAS maps (Schlegel et al. 1998).  Finally, we adopt
$\sigma_{R_V}=0.3$. Note that for typical values of  $E(B-V) \sim
0.1$, $\sigma_{m_{\lambda}}$  amounts to  $\sim 0.05$ magnitudes at
$V$ and $I$ wavelengths.

\section{The Calibration of $K_s$ and $K'$ SBF}

SBF measurements in the $K'$ and $K_s$ bands do not comprise a
sufficiently large sample size or have the required accuracy to
determine the dependence of the SBF absolute magnitude on metallicity.
Zero points have been determined using galaxies in common with $I$-SBF
surveys by Pahre \& Mould (1994) as $\overline{M}_{K_s}=-5.74 \pm
0.18$ (based on a sample of seven Virgo galaxies, plus NGC 3379, M31
and M32, the latter two observed in the $K'$ band) and Jensen, Tonry \& Luppino
(1998) as $\overline{M}_{K'}=-5.61\pm0.12$ (based on a sample
of Virgo, Fornax and Eridanus galaxies, plus M31 and M32).

Because a satisfactory overlap is not present with Cepheid galaxies,
the calibration of both $K_s$ and $K'$ SBF magnitudes must proceed
through $I$-SBF. Assuming the calibration given in equation (5)  for
$I$-SBF, we derive $\overline{M}_{K_s}=-6.01 \pm 0.11$  and
$\overline{M}_{K'}=-5.77 \pm 0.11$  using DIRBE/IRAS reddenings, and
assuming no metallicity dependence of the Cepheid PL relation.  NGC
4489 shows large residuals from the mean in $K_s$  datasets, and has
therefore been rejected from the fit.   Following Jensen et
al. (1998), we have also rejected NGC 1339, NGC 1395, NGC 1400, NGC
1426, NGC 4489 and NGC 4578 from the calibration of $K'$.  The
difference between our calibration and the ones from Pahre \& Mould
(1994) and Jensen, Tonry \& Luppino (1998) are due  to the different
magnitude zero point adopted for $I$-SBF, the switch to DIRBE
reddenings, and revision to the published $I$ fluctuation magnitudes
(Tonry et al. 1999). 

We must stress  that a final calibration of the near-infrared SBF must
await a larger dataset (Jensen et al. 1998), in particular
the dependence of the fluctuation magnitude on color should be empirically
determined. Figure 13 shows the color dependence
of the fluctuation magnitudes given the available data. Formal fits
give slopes of $-8.3\pm4.6$ in $K_s$ and $+5.0\pm0.8$ in $K'$, however
it is rather obvious that these estimates cannot be reliable in view
of the fact that the slopes are expected to be similar in the two
passbands.

\section{Magnitude Zero Points Using HI Reddenings}

Magnitude zero points for TRGB, PNLF, GCLF and SBF, derived as in \S
3$-$\S6, but  using HI rather than DIRBE/IRAS reddenings, are
given in Table 6. In no cases are the differences significant.
Note that, because HI reddenings are systematically smaller than the
DIRBE/IRAS reddenings, the effect of adopting them is to make the zero
points for PNLF, TRGB and GCLF fainter by a few hundredth of a
magnitude. The effect on SBF is opposite, i.e. the zero points become
brighter, due to the color dependence of the SBF calibration (\S
6).

\clearpage

\begin{figure}
\figurenum{1}
\plotone{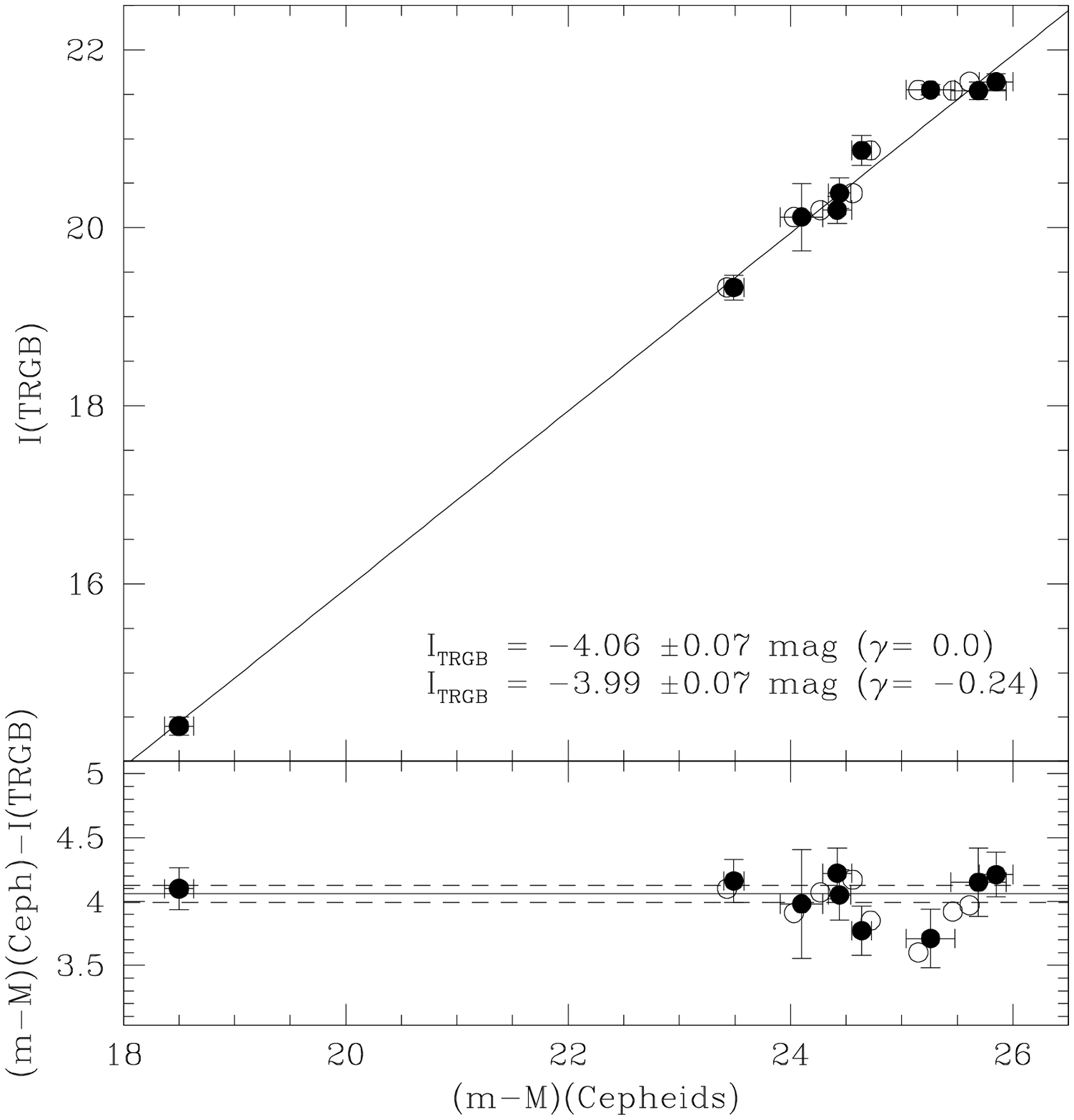}
\caption{Zero point calibration for the TRGB method, based on a galaxy
by galaxy comparison. In this and all subsequent figures, all
magnitudes have been  corrected for extinction using DIRBE/IRAS
reddenings and the extinction coefficients given in \S 2.  Filled
circles assume no dependence of the PL relation on metallicity: the
solid line in the upper panel shows  the best fitting line, assuming a
fixed slope of 1.0, and determined considering errors in both
variables.   The same line, and the 1$\sigma$ uncertainty of the mean, 
is shown in the lower panel, where residuals are plotted.
Small open circles assume a metallicity dependence as in Kennicutt et
al. (1998, $\gamma = \delta(m-M)_0/\delta[O/H] = -0.24$, the error bars
are comparable to the previous case).  The zero points and 1$\sigma$
errors are listed in the figure for both of the assumed metallicity
dependences. The galaxies are, from left to right: LMC, NGC 6822,
IC 10, IC 1613, M31, M33, NGC 3109, Sextans B and Sextans A. }

\end{figure}

\clearpage

\begin{figure}
\figurenum{2} 
\plotone{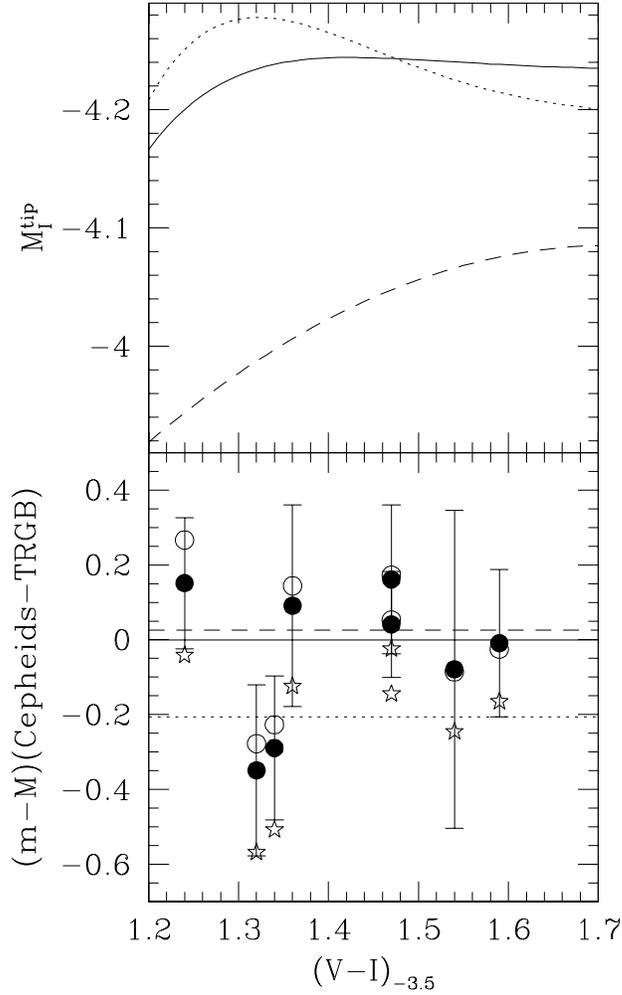}
\caption{The top panel shows the color dependence of the absolute $I$
magnitude of the TRGB predicted by Lee et al. (1993, dashed line) and
Salaris and Cassisi (1998, the dotted and solid lines use Kurucz'
stellar models and  Yale isochrones respectively).  The lower panel
shows the Cepheid $-$ TRGB distance moduli residuals using:  the
calibration given in  equation (1) (solid dots and error bars); the
Lee et al. calibration (open circles, with a mean shown by the dashed
line); the Salaris \& Cassisi calibration (stars, with a mean given by
the dotted line). The galaxies are, from left to right: Sextans A, NGC
3109, NGC 598, Sextans B, LMC, IC 1613, IC 10, and NGC 224. NGC 6822
is not plotted since its $(V-I)_{-3.5}$ color places it outside the
range of metallicities considered by the primary calibration. }
\end{figure}

\clearpage

\begin{figure}
\figurenum{3} 
\plotone{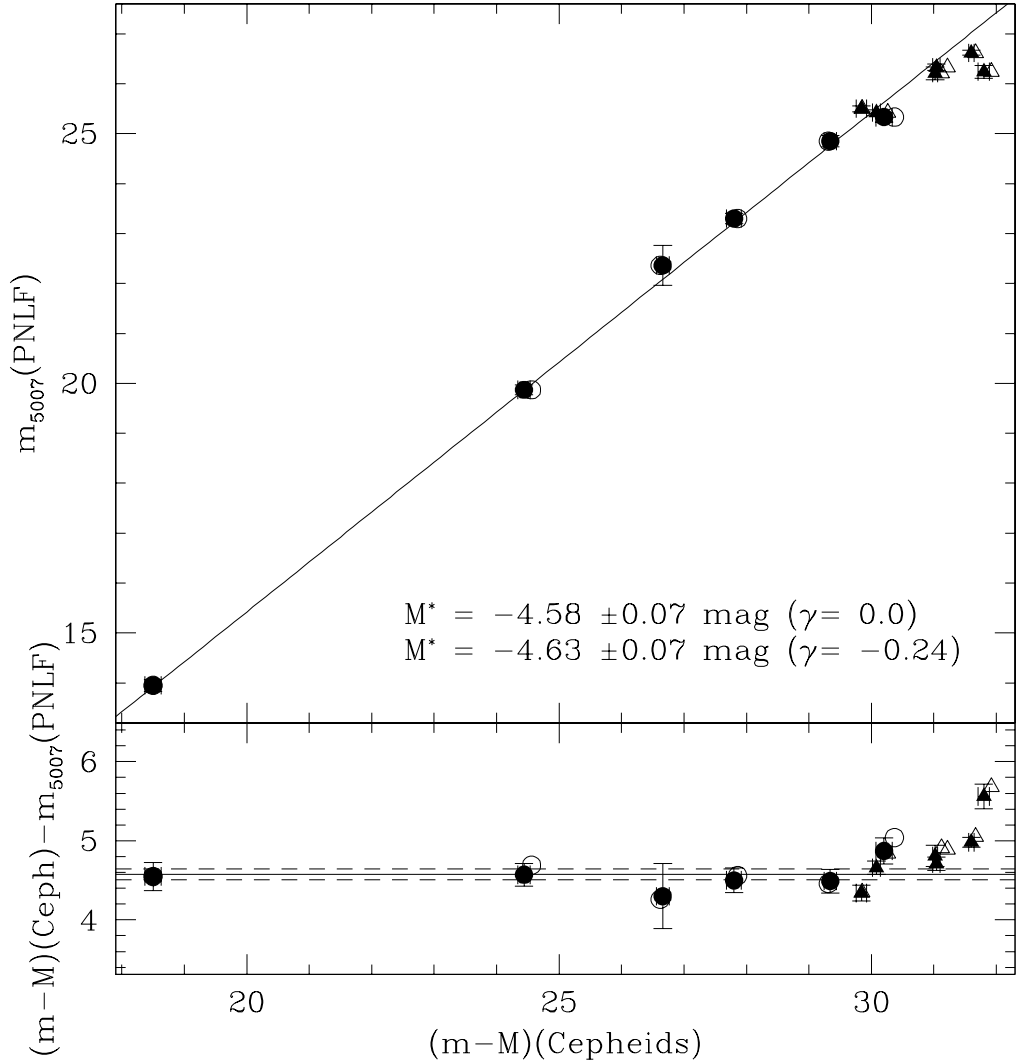}
\caption{Direct zero point calibration for the planetary nebula
luminosity function method, based on galaxy comparison with the
Cepheids. Shown in the figure are the calibrator galaxies (circles)
and also for comparison  group distances and mean PNLF cutoff
magnitudes (triangles). The filled and open symbols refer to the
degree of metallicity dependence as in Figure 1.  The galaxies are,
from left to right:  LMC, M31,  NGC 300, M81, M101 and NGC 3368. The
groups are: NGC 1023 Group, the Leo I group, the NGC 4472 subcluster,
the M87 subcluster, Fornax and the NGC 4649 subcluster. The fits
leading to the magnitude zero points listed in the figure are
performed using only the six galaxies.}
\end{figure}

\clearpage

\begin{figure}
\figurenum{4}
\plotone{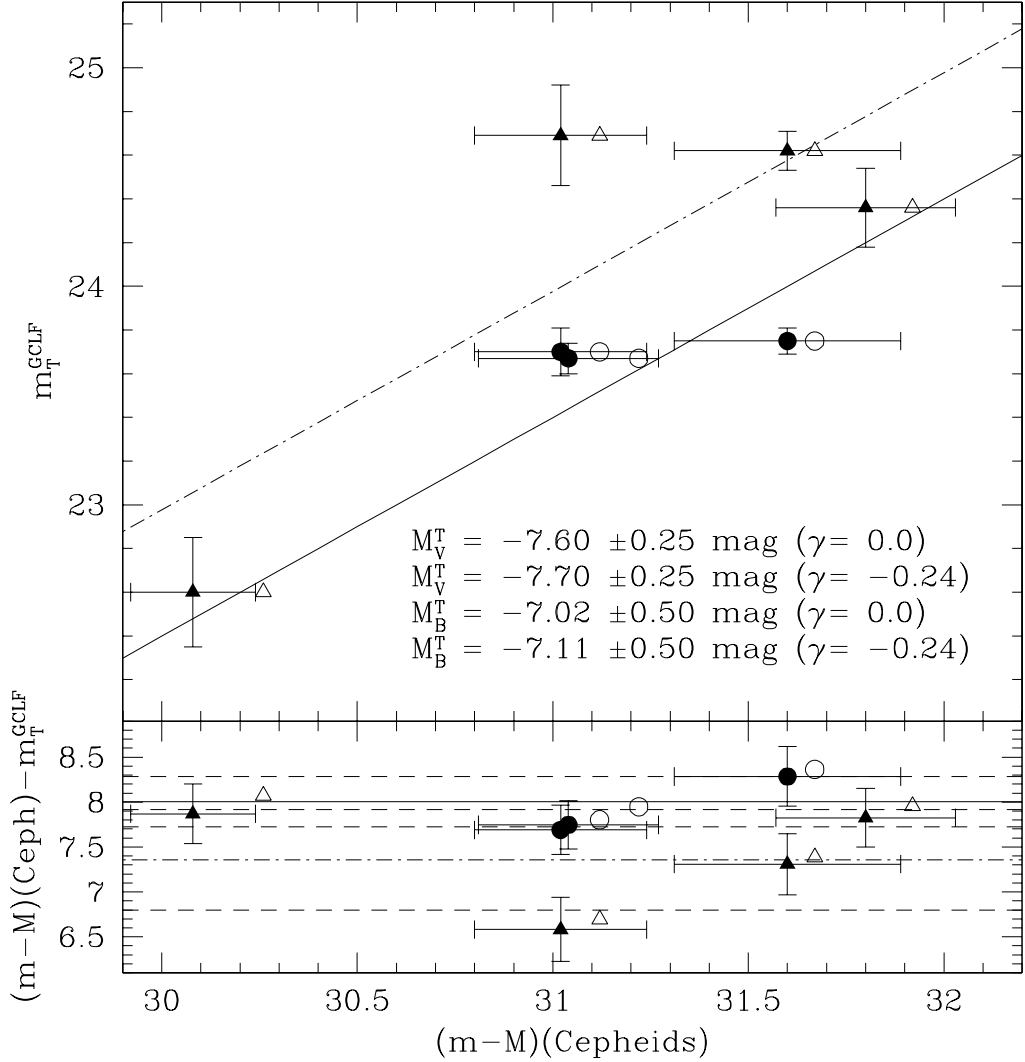}
\caption{Zero point calibration for the $V-$band (circles) and
B-band (triangles) GCLF methods, based on a group by group
comparison with the Cepheids. Filled and open symbols refer to the
degree of metallicity dependence of the Cepheid PL relation, as in
Figure 1. The groups are, from left to right: $V-$band GCLF: the NGC
4472 subcluster,  the M87 subcluster, and Fornax cluster;
$B-$band GCLF: Leo I group, the NGC 4472  subcluster, Fornax and
the NGC 4649 subcluster. Formal fits are shown by the solid line
for the $V-$band GCLF and by the dot-dashed line for the $B-$band
GCLF.}

\end{figure}

\clearpage

\begin{figure}
\figurenum{5}
\plotone{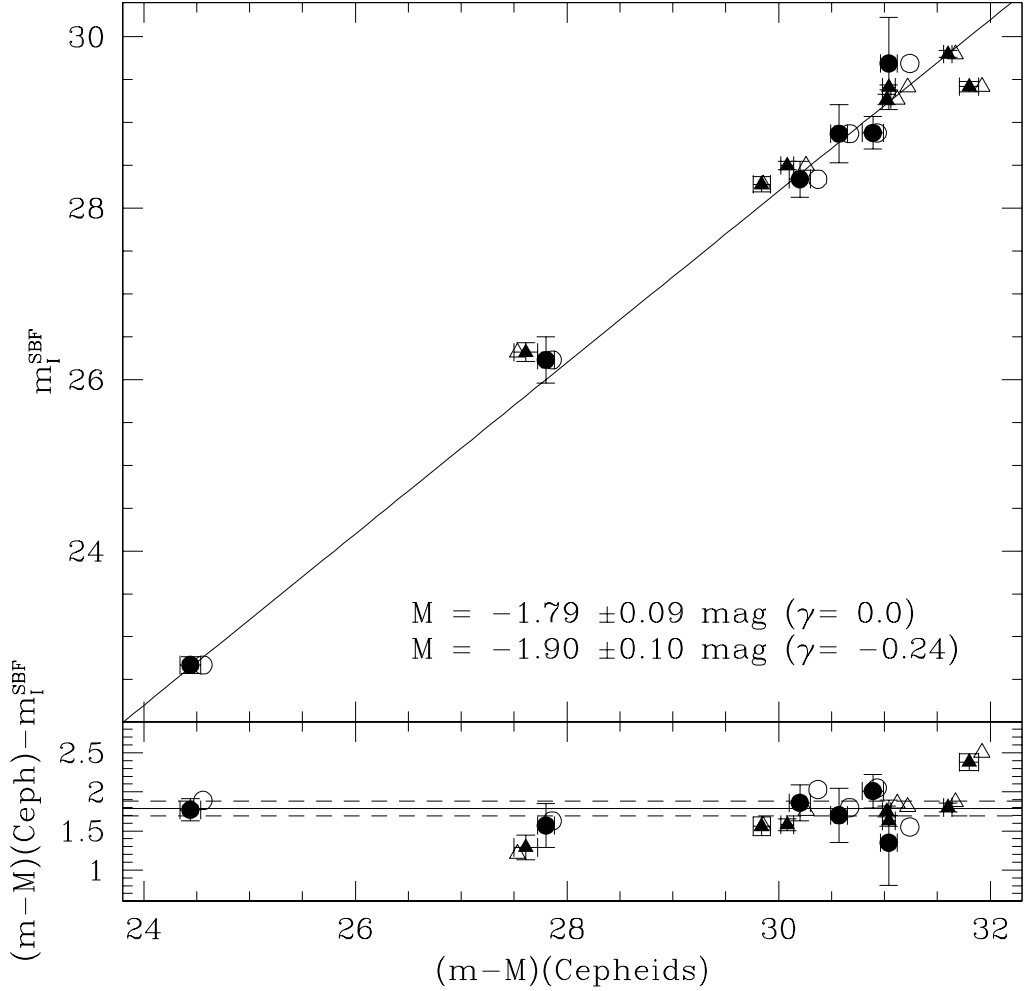}
\caption{Zero point calibration for the $I$-SBF method, based on a
direct comparison of galaxies with Cepheid distances and SBF
fluctuation magnitudes. A color correction has been applied to the SBF
magnitudes as described in \S 6. For comparison, both galaxies
(circles) and clusters (triangles) are plotted in the figure.  The
filled and open symbols refer to the degree of metallicity dependence
as in Figure 1. The galaxies are, from left to right: M31, M81, NGC
3368, NGC 4725, NGC 7331 and NGC 4548. The groups are the NGC 5128
group, NGC 1023 group, Leo I group, the M87 and NGC 4472
subclusters, Fornax cluster and the NGC 4649 subcluster.}
\end{figure}

\clearpage

\begin{figure}
\figurenum{6}
\plotone{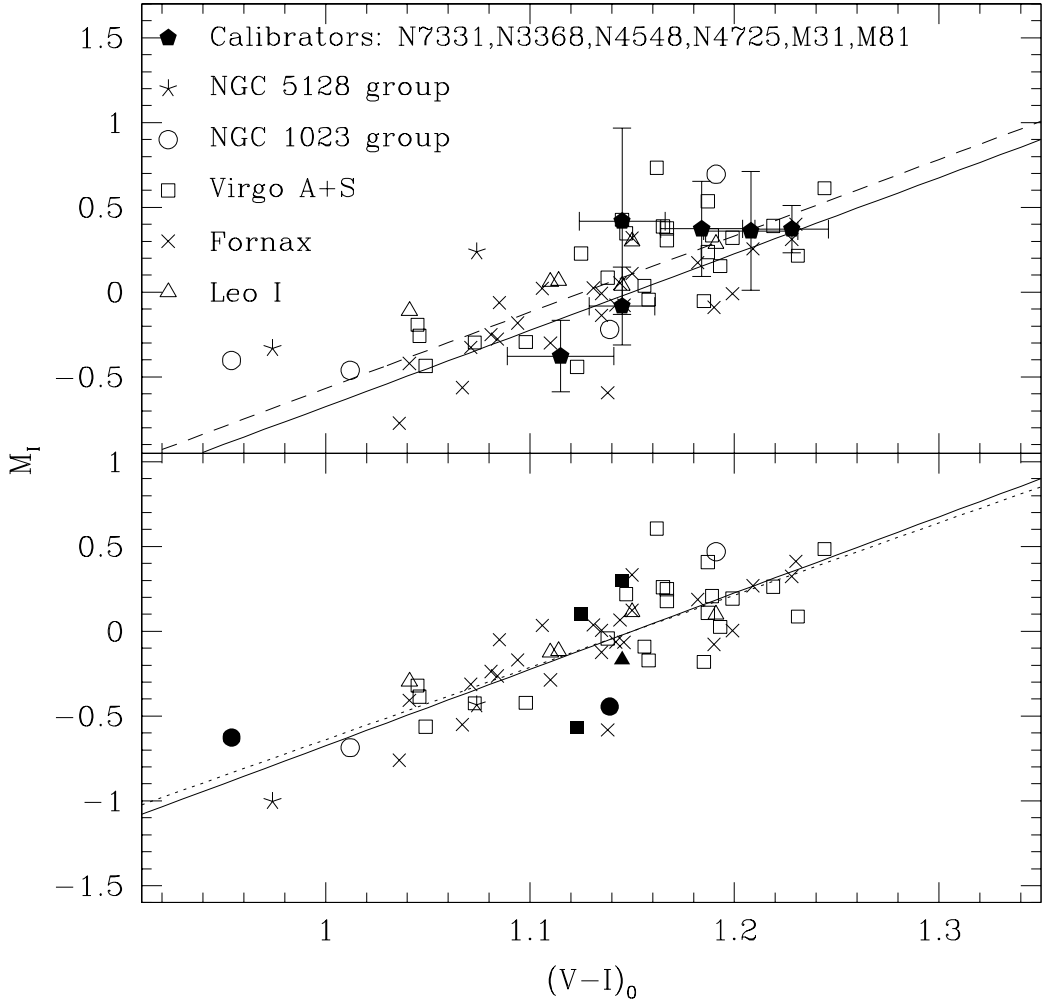}
\caption{In the lower panel are plotted galaxies belonging to the
groups identified in the top left, shifted vertically according to
their mean magnitude at $(V-I)_0=1.15$, following Tonry et
al. (1997). The open and filled symbols refer to elliptical and spiral
galaxies respectively. The solid line has a slope of 4.5, as in Tonry
et al. (1997), while the dotted line is our best fit to the data. In
the top panel  the galaxies are shifted vertically according to the
mean Cepheid distance of the group to which they belong. The Cepheid
galaxies on which our SBF calibration (shown by the solid line) is
based are shown by the solid pentagons. A calibration based
exclusively on group distances is shown by the dashed line.}
\end{figure}

\clearpage

\begin{figure}
\figurenum{7} \plotone{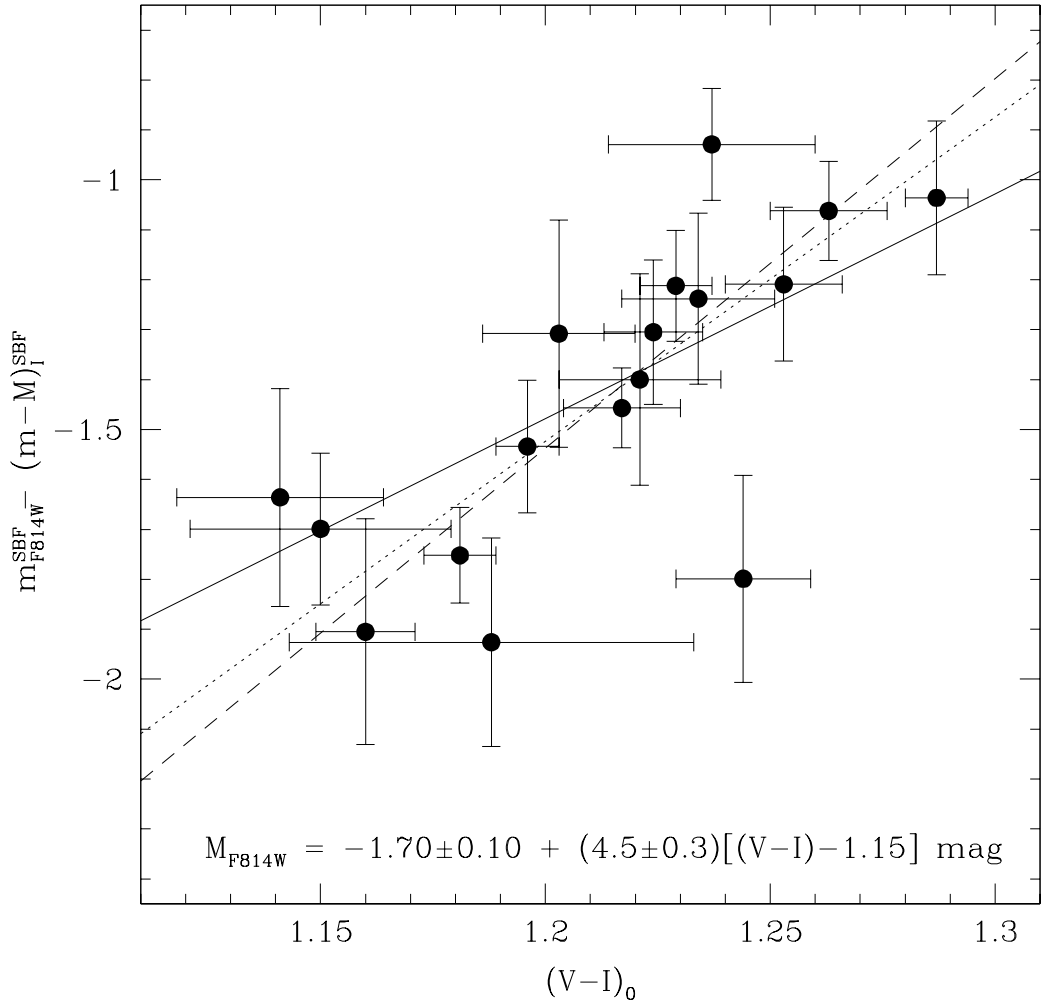}
\caption{Tertiary  calibration for the F814W-SBF method, calibrated
against the $I$-SBF method.  The lines are as follows. Solid line: our
adopted best fit which assumes a 4.5 slope, as in equation (6); dotted
line: fit with an assumed slope of 6.5 as in Ajhar et al. (1997); dashed
line:  unconstrained fit (both slope and zero point). M31 is the point
at (1.237,$-$0.936).}
\end{figure}

\clearpage

\begin{figure}
\figurenum{8}
\plotone{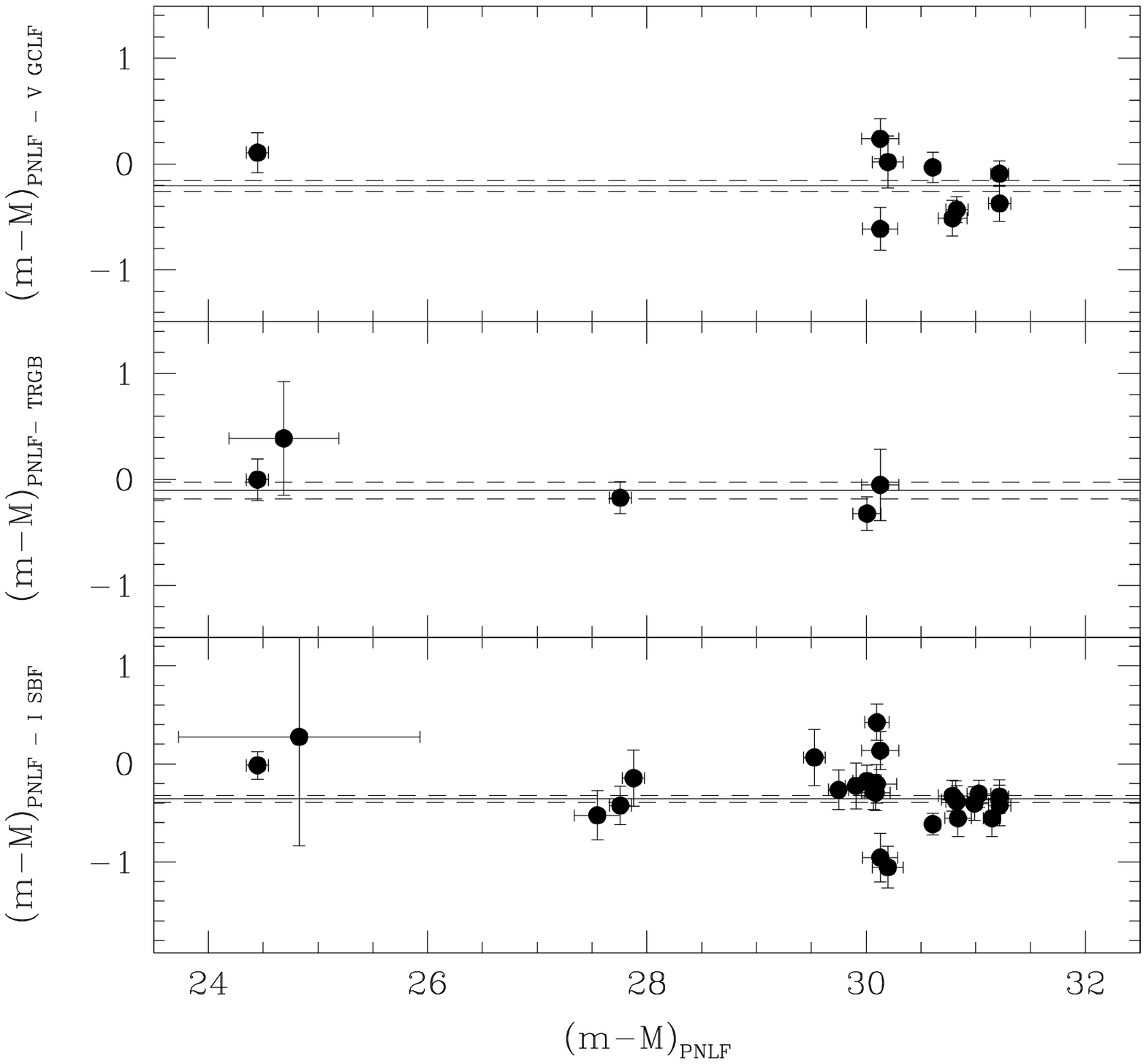}
\caption{Comparison between galaxy distance moduli derived using the
PNLF method  and the $I$-SBF, TRGB and $V-$GCLF methods, calibrated as
in Table 1. The solid and dashed lines are the best fit to the galaxy data
points and the 1$\sigma$ deviations respectively.}
\end{figure}

\clearpage

\begin{figure}
\figurenum{9}
\plotone{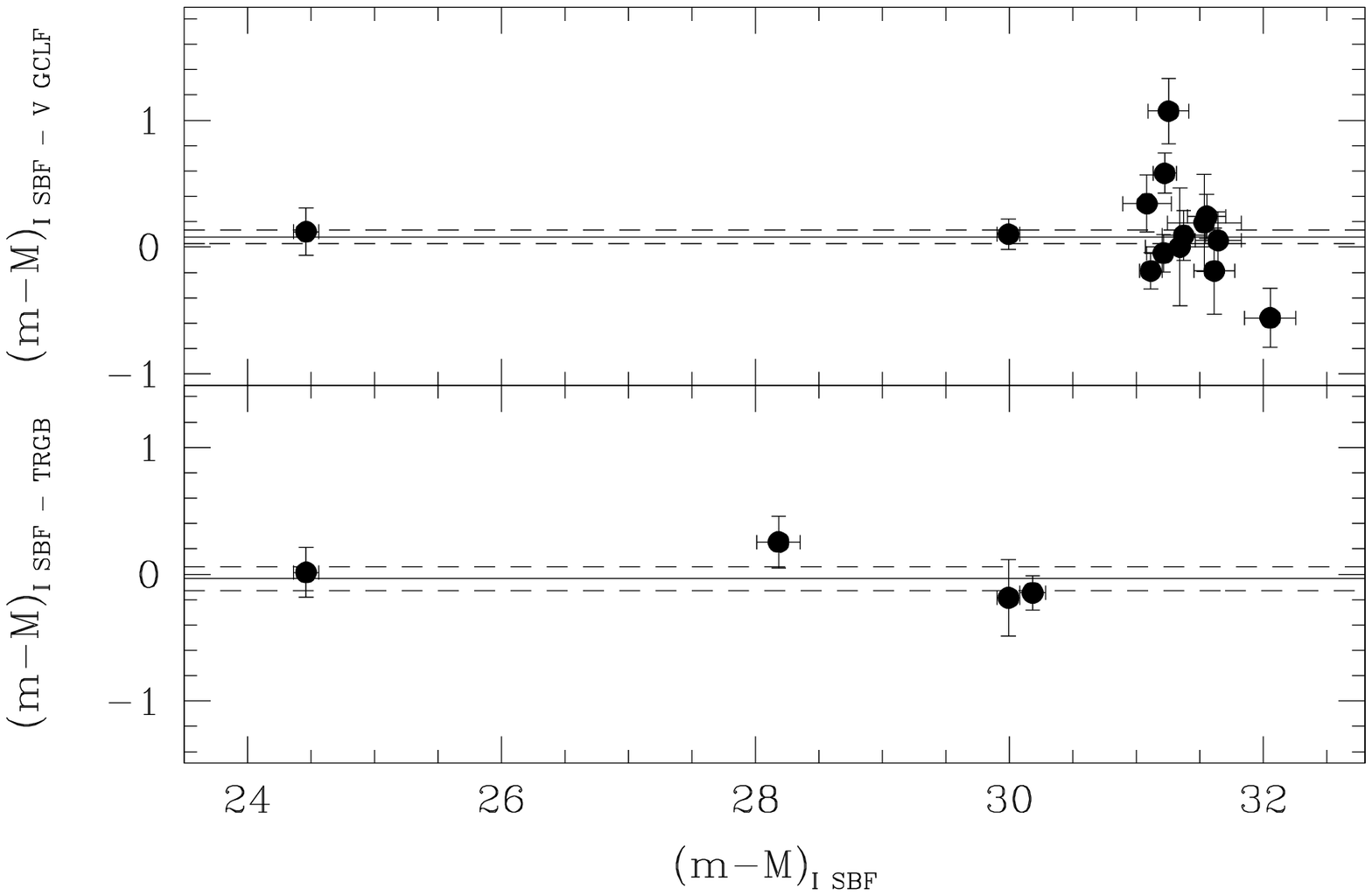}
\caption{Comparison between galaxy distance moduli derived using the
$I$-SBF method  and the TRGB and $V-$GCLF methods, calibrated as
in Table 1. The solid and dashed lines are the best fit to the galaxy data
points and the 1$\sigma$ deviations respectively.}
\end{figure}

\clearpage

\begin{figure}
\figurenum{10}
\plotone{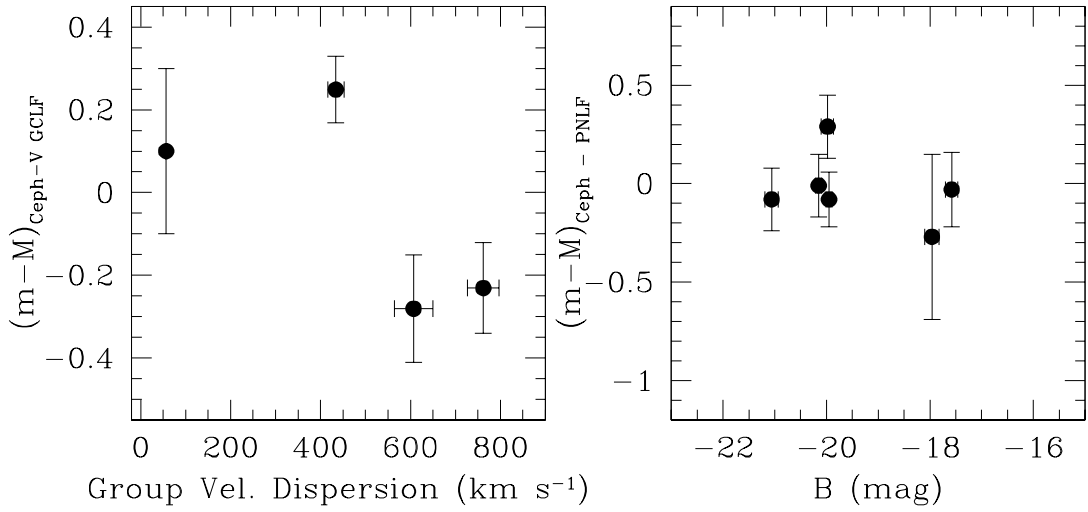}
\caption{Cepheid$-V$ GCLF and Cepheid$-$PNLF distance moduli
residuals plotted against  the cluster velocity dispersion and of the
absolute $B$ magnitude of the host galaxy respectively. }
\end{figure}

\clearpage

\begin{figure}
\figurenum{11}
\plotone{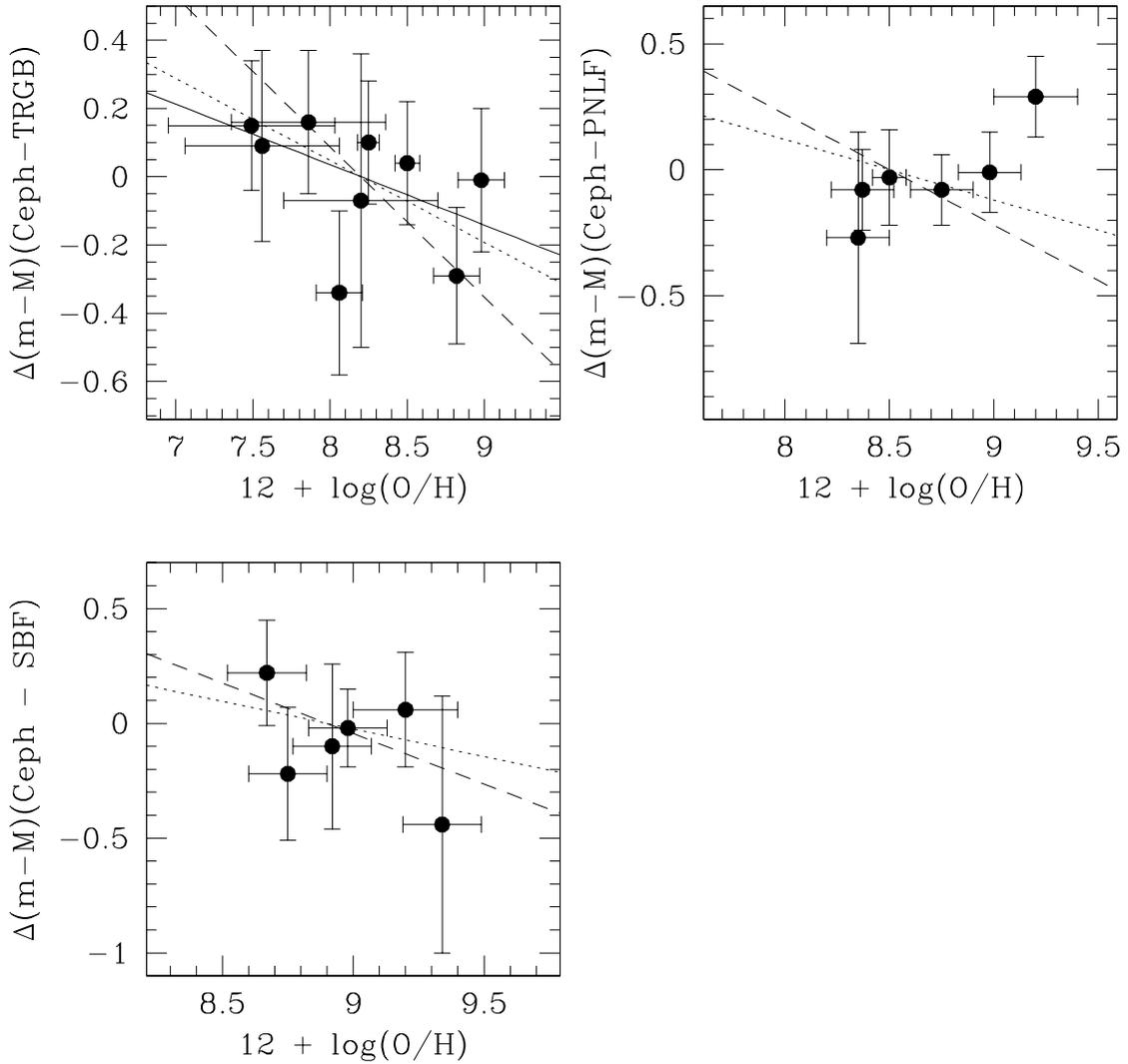}
\caption{Test for the metallicity dependence of the Cepheid PL
relation. Cepheids abundances are plotted in the x-axis.
The solid line is a bivariate fit to the data points. The
dotted lines show the relation expected based on the metallicity
dependence of the Cepheid PL relation derived by  Kennicutt et
al. (1998). The dashed lines show the relation expected given the
metallicity  dependence of Sasselov et al. (1997) and Beaulieu et
al. (1997).}
\end{figure}

\clearpage

\begin{figure}
\figurenum{12}
\plotone{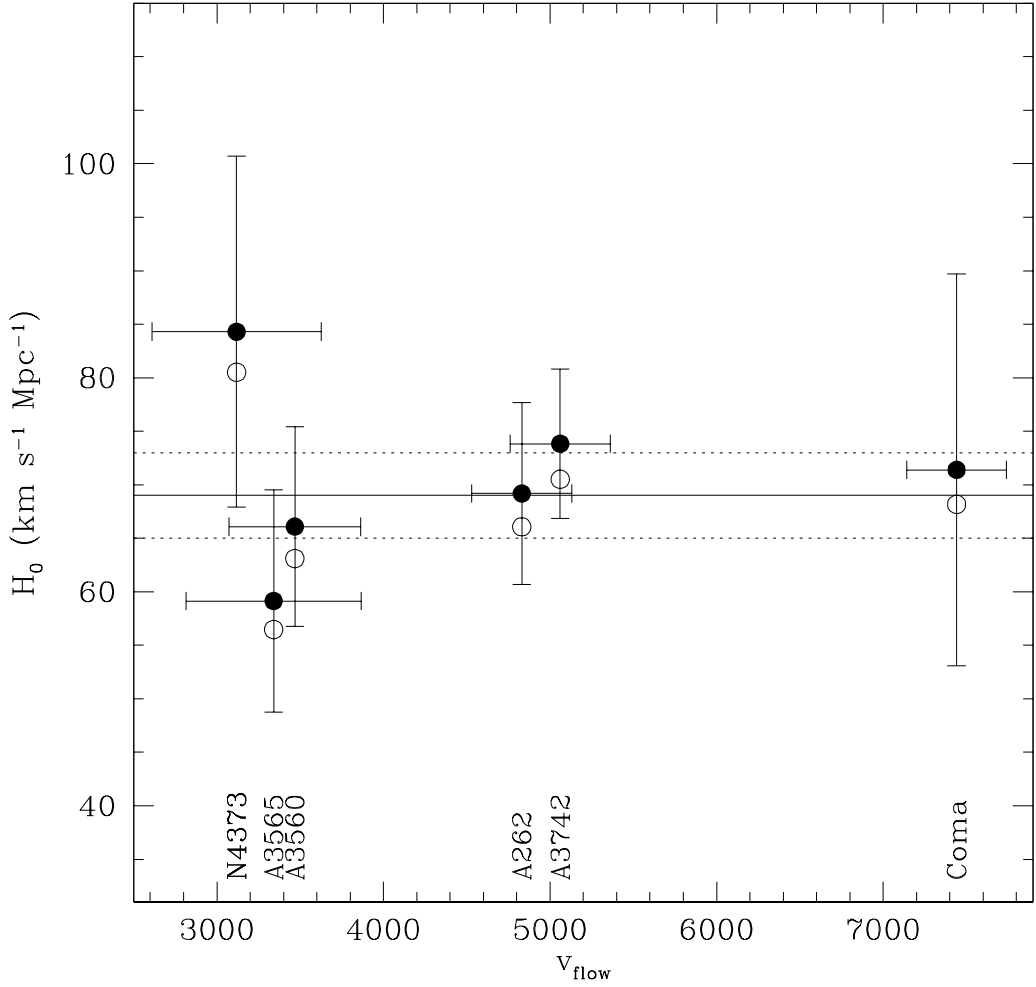}
\caption{Values of the Hubble constant, $H_0$, derived from
$HST$/WFPC2/F814W-SBF distances, calibrated as described in \S
6. Group velocities are corrected for a flow model (see text for
further details).   The solid and open points assume no metallicity
dependence for the Cepheid PL relation, and a metallicity dependence
as in Kennicutt et al. (1998) respectively. The solid and dotted lines
give the weighted mean and 1$\sigma$ error of the values of $H_0$
derived for the four Abell clusters, assuming no metallicity
dependence of the Cepheid PL relation.}
\end{figure}

\clearpage

\begin{figure}
\figurenum{13}
\plotone{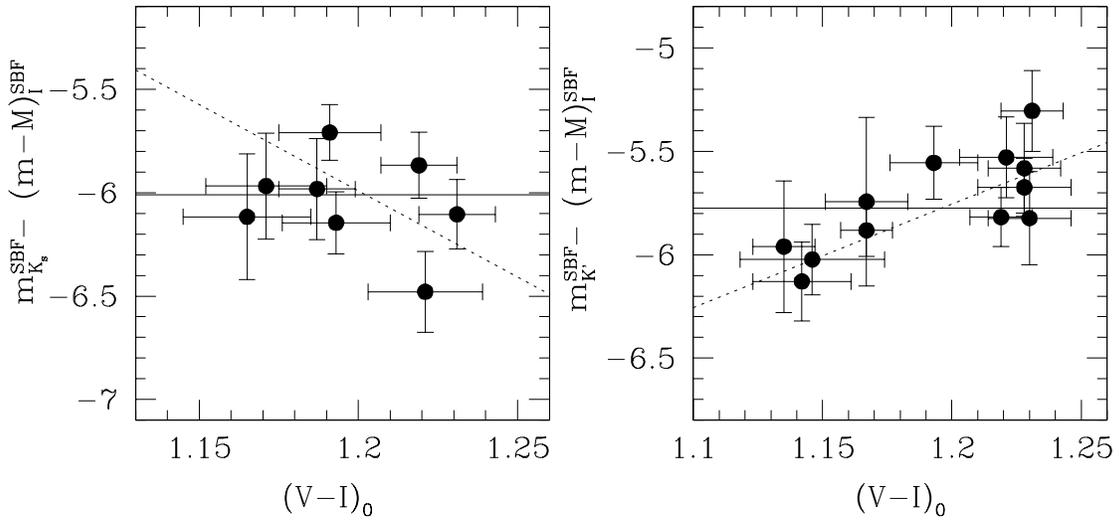}
\caption{Color dependence of the near infrared SBF method. Formal bivariate fits
to the data give slopes of $-8.3\pm4.6$ in $K_s$ and $5.0\pm0.8$ in $K'$, shown by the dotted lines.
Our adopted zero points, assuming no color dependence, are shown by the solid lines.}
\end{figure}

\begin{deluxetable}{llllll}
\tablecolumns{6}
\tablewidth{0pc}
\tablenum{1}
\tablecaption{Calibration of Secondary Distance Indicators\label{tbl-1}}
\tablehead{
\colhead{Indicator} &
\colhead{No.} &
\colhead{Gal/} &
\colhead{Zero Point\tablenotemark{2}} &
\colhead{Zero Point\tablenotemark{3}} &
\colhead{Notes} \nl
\colhead{} &
\colhead{} &
\colhead{Group\tablenotemark{1}} &
\colhead{$\gamma$=0.00} &
\colhead{$\gamma= -0.24 \pm 0.16$} &
\colhead{} 
}
\startdata
TRGB   & 9 & Gal & \underline{$-$4.06$\pm$0.07} (0.80) & $-$3.99$\pm$0.07 (0.63)   &  \nl
       & 4 & Gr  & $-$4.00$\pm$0.05 (4.03) & $-$4.11$\pm$0.06 (6.07)               &  \nl
PNLF   & 6 & Gal & \underline{$-$4.58$\pm$0.07} (0.83) & $-$4.63$\pm$0.07 (1.60)               &   \nl  
       &10 & Gr+Gal&$-$4.61$\pm$0.06 (0.83) & $-$4.68$\pm$0.06 (1.40)  & \nl  
V GCLF & 3 & Gr  & \underline{$-$7.60$\pm$0.25} (12.0) & $-$7.70$\pm$0.25 (7.98)   &  \nl  
       & 1 & Gal & \underline{$-$7.70$\pm$0.19} (N/A) & $-$7.82$\pm$0.19 (N/A)   & 4  \nl  
B GCLF & 4 & Gr  & \underline{$-$7.02$\pm$0.50} (5.48) & $-$7.11$\pm$0.50 (6.03)   & \nl  
I SBF  & 6 & Gal &  \underline{$-$1.79$\pm$0.09} (0.51) & $-$1.90$\pm$0.10 (0.43)               &  5 \nl
       & 6 & Gr  & $-$1.69$\pm$0.03 (2.87) & $-$1.79$\pm$0.03 (3.25)               & 5  \nl
F814W-SBF &17&Gal& \underline{$-$1.70$\pm$0.10} (1.38) & $-$1.81$\pm$0.10 (1.38) & 6,7 \nl  
            &4&Gr& $-$1.50$\pm$0.04 (4.19) & $-$1.63$\pm$0.04 (6.70) & 7 \nl  
\enddata
\tablenotetext{1}{`Gal' denotes a direct calibration based on a 
galaxy-by-galaxy comparison, `Gr' denotes a calibration based on group distances, `Gal+Gr' denotes a mixed calibration based both on group and galaxy distances. Unless noted, all calibrations are based on Cepheid distances. Details for each indicators can be found in the text and accompanying figures. Cluster depth effects are included in the errors on the group distances for the mixed calibration,
but not for the calibration based exclusively on groups.}
\tablenotetext{2}{Derived assuming no metallicity dependence of the
Cepheid PL relation. The error is the $1\sigma$ uncertainty in the
fit, see \S 3$-$6 for an estimate of the systematic uncertainties. The
numbers in parentheses represent the reduced $\chi^2$ of the fit. Our
adopted final magnitude zero points are underlined in the table.}
\tablenotetext{3}{Derived assuming a metallicity dependence of the Cepheid PL
relation as in Kennicutt et al. (1998).}
\tablenotetext{4}{Based on a direct comparison for M31 only.}
\tablenotetext{5}{The calibration assumes a color dependence of the fluctuation magnitudes as in equation (5).}
\tablenotetext{6}{Calibrated against the $I$-SBF distances, calibrated as in equation (5).}
\tablenotetext{7}{The calibration assumes a color dependence as in equation (6).}
\end{deluxetable}

\clearpage

TABLE 2 IS IN LANDSCAPE FORMAT, THEREFORE IN A SEPARATE FILE

\begin{deluxetable}{ll}
\tablecolumns{2}
\tablewidth{0pc}
\tablenum{3}
\tablecaption{Comparisons of Secondary Distance Indicators\label{tbl-1}}
\tablehead{
\colhead{Indicators}&
\colhead{$\Delta \pm \sigma$Difference\tablenotemark{a}}\nl
\colhead{}&
\colhead{(mag)}
}
\startdata
PNLF $-$ TRGB        &$-$0.12 $\pm$ 0.08 \nl
PNLF $-$ SBF         &$-$0.36 $\pm$ 0.04 \nl  
PNLF $-$ $V$ GCLF    &$-$0.22 $\pm$ 0.05 \nl  
$I$ SBF $-$ TRGB     &$-$0.03 $\pm$ 0.09 \nl  
$I$ SBF $-$ $V$ GCLF &+0.08 $\pm$ 0.05   \nl  
\enddata
\tablenotetext{a}{Weighted mean of the difference in the distance moduli derived
using galaxies in common between the two distance indicators listed in the first column.
The sigma represents the rms error in the mean.}
\end{deluxetable}

\clearpage

TABLE 4 IS IN LANDSCAPE FORMAT, THEREFORE IN A SEPARATE FILE

\clearpage

\begin{deluxetable}{llll}
\tablecolumns{4}
\tablewidth{0pc}
\tablenum{5}
\tablecaption{Error Budget \label{tbl-10}}
\tablehead{
\colhead{ } &
\colhead{Source} &
\colhead{Error\tablenotemark{1}} &
\colhead{Notes} 
}
\startdata
{\bf 1.} & \multicolumn{3}{l}{\bf  ERRORS ON THE CEPHEID DISTANCE SCALE} \nl
   &{\it A. LMC True Modulus} &  $\pm$0.13 &  \nl 
   &{\it B. LMC PL Zero Point \tablenotemark{2}} & $\pm$0.02 & \nl
S$_{1.1}$ & LMC PL Systematic Error& $\pm$0.13&A and B added in quadrature\nl
   &{\it C. $HST$ V-Band Zero Point\tablenotemark{3}} & $\pm$0.03&  \nl
   &{\it D. $HST$ I-Band Zero Point\tablenotemark{3}} & $\pm$0.03&  \nl
S$_{1.2}$ & Systematic Error in the Photometry& $\pm$0.09 & $\sqrt{C^2(1-{\rm R})^2+D^2{\rm R}^2}$, R$=A(V)/E(V-I)$ \nl
R$_{1.1}$ & Random Error in the Photometry & $\pm$0.05 & From DoPHOT/ALLFRAME comparison  \nl
   &{\it E. $R_V$ Differences Between Galaxy and LMC} & $\pm$0.014 &See Ferrarese et al. (1998) for details\nl
   &{\it F. Error in the adopted value for R$_V$} & $\pm$0.01 & See Appendix A for details\nl 
R$_{1.2}$ & Random Error in the Extinction Treatment& $\pm$0.02 & E and F added in quadrature \nl
   &{\it G. PL Fit (V)} &  $\pm$0.05\tablenotemark{4} &  \nl
   &{\it H. PL Fit (I)} &  $\pm$0.04\tablenotemark{4} &  \nl
R$_{1.3}$ & Random Error in the Cepheid True Modulus\tablenotemark{5} & $\pm$0.06\tablenotemark{4} & G and H partially correlated.\nl
R$_{PL}$  & Total Random Error&{\bf $\pm$0.08}\tablenotemark{4}  &R$_{1.1}$, R$_{1.2}$ and R$_{1.3}$ added in quadrature \nl
S$_{PL}$  & Total Systematic Error&{\bf $\pm$0.16} & S$_{1.1}$ and S$_{1.2}$ added in quadrature \nl
& & \nl
{\bf 2.} & \multicolumn{3}{l}{\bf ERRORS ON THE F814W-SBF DISTANCE SCALE} \nl
   &{\it A. Error on the Reddened SBF Magnitudes} & $\pm$0.10\tablenotemark{4}& see F99 \nl
   &{\it B. Error on the Reddened Colors}                 & $\pm$0.02\tablenotemark{4}& see F99 \nl
   &{\it C. Error on the $A$(F814W) Extinction}       & $\pm$0.03\tablenotemark{4}   &  See Appendix A  for details  \nl
   &{\it D. Error on the $E(V-I)$ reddening}           & $\pm$0.01\tablenotemark{4}   & See Appendix A  for details   \nl
R$_{2.1}$ & Random Error on the SBF Magnitudes & $\pm$0.10\tablenotemark{4} & A and C added in quadrature \nl
R$_{2.2}$ & Random Error on the Colors & $\pm$0.02\tablenotemark{4} & B and D added in quadrature \nl
R$_{2.3}$ & Random Error on the Color Corrected Mag. & $\pm$0.15\tablenotemark{4} & $\sqrt{R_{2.1}^2+(4.5R_{2.2})^2+(0.3((V-I)_0-1.15))^2}$ \nl
S$_{2.1}$ & Systematic Error on the Zero Point  & $\pm$0.10 & From Table 1, R$_{PL}$ already folded in  \nl
R$_{SBF}$  & Total Random Error&{\bf $\pm$0.15}\tablenotemark{4} &R$_{2.3}$ \nl
S$_{SBF}$  & Total Systematic Error&{\bf $\pm$0.19}\tablenotemark{4} & S$_{2.1}$ and S$_{PL}$ added in quadrature \nl
& & \nl
{\bf 3.} &\multicolumn{3}{l}{\bf ERRORS ON $H_0$}  \nl
R$_{3.1}$ &Random Error on the Flow Velocities  & $\pm$400\tablenotemark{4} & in \kms \nl
R$_{H_0}$  & Total Random Error on $H_0$ (km/s/Mpc)&{\bf $\pm$4} & $\sqrt{(R_{3.1}/d)^2 + (0.46H_0R_{SBF})^2}/\sqrt N$\nl
S$_{H_0}$  & Total Systematic Error on $H_0$ (km/s/Mpc)&{\bf $\pm$6} &  $0.46H_0S_{SBF} $\nl
\enddata
\end{deluxetable}

\clearpage

\noindent$^{1}${Errors in 1. and 2. are in magnitudes, and as indicated in
3.} 

\noindent$^{2}${Equal to the scatter around the dereddened
PL relation for the LMC ($\pm$0.12 mag) divided by the square root
of the number of LMC Cepheids, 32 (Madore \& Freedman 1991).}

\noindent$^{3}${Contributing uncertainties from 
the Holtzman et al. (1995) zero points, and the `long versus short' uncertainty,
combined in quadrature.}

\noindent$^{4}${The values quotes are typical of the galaxies observed, but 
individual cases vary slightly. See the individual references for the
correct values of specific galaxies.}

\noindent$^{5}${The partially correlated nature of the derived PL
width uncertainties  is taken into account by the (correlated)
de-reddening procedure, coupled  with the largely
`degenerate-with-reddening' positioning of individual  Cepheids within
the instability strip.}

\clearpage

\begin{deluxetable}{llcl}
\tablecolumns{4}
\tablewidth{0pc}
\tablenum{6}
\tablecaption{Final Zero Points Using HI Reddenings\label{tbl-1}}
\tablehead{
\colhead{Indicator} &
\colhead{Zero Point\tablenotemark{1}} &
\colhead{$\Delta$\tablenotemark{2}} &
\colhead{Notes} 
}
\startdata
TRGB     & $-$4.03$\pm$0.07 (0.72) & $-$0.02 & Galaxy calibration via Cepheids \nl
PNLF     & $-$4.53$\pm$0.07 (1.00) & $-$0.05 & Galaxy calibration via Cepheids \nl
$V$ GCLF & $-$7.57$\pm$0.25 (10.8) & $-$0.03 & Group calibration via Cepheids \nl
$B$ GCLF & $-$6.96$\pm$0.50 (5.50) & $-$0.06 & Group calibration via Cepheids \nl
$I$-SBF  & $-$1.84$\pm$0.10 (0.41) & +0.04 & Galaxy calibration via Cepheids \nl
F814W-SBF& $-$1.73$\pm$0.11 (0.99) & +0.02 & Galaxy calibration via $I$-SBF, calibrated as above\nl
$K'$-SBF & $-$5.75:$\pm$0.11 (1.85) & $-$0.03 & as above   \nl  
$K_s$-SBF& $-$6.01:$\pm$0.13 (1.54) & $-$0.01 & as above, excluding NGC 4489 \nl
\enddata
\tablenotetext{1}{Derived assuming no metallicity dependence of the
Cepheid PL relation. The error is the $1\sigma$ uncertainty in the
fit, see \S 3$-$6 for an estimate of the systematic uncertainties. The
numbers in parentheses represent the reduced $\chi^2$ of the fit.}
\tablenotetext{2}{Difference between the magnitude zero points derived
using DIRBE/IRAS (from Table 1) and HI reddenings.}
\end{deluxetable}

\end{document}


\begin{deluxetable}{lcllllllllll}                        
\tablecolumns{12}                        
\tablewidth{0pc}                        
\scriptsize        
\tablenum{2}                
\tablecaption{Final Cepheid, TRGB, PNLF, GCLF, \& SBF Distance Moduli\tablenotemark{1}\label{tbl-2}}                    
\tablehead{                        
\colhead{Galaxy} &                       
\colhead{Class\tablenotemark{2}} &                       
\colhead{$(m-M)$\pp$\sigma$} &                      
\colhead{$(m-M)$\pp$\sigma$} &                      
\colhead{$(m-M)$\pp$\sigma$} &                      
\colhead{$(m-M)$\pp$\sigma$} &                      
\colhead{$(m-M)$\pp$\sigma$} &                      
\colhead{$(m-M)$\pp$\sigma$} &                      
\colhead{$(m-M)$\pp$\sigma$} &                      
\colhead{$(m-M)$\pp$\sigma$} &                      
\colhead{$(m-M)$\pp$\sigma$} &                      
\colhead{$(m-M)$\pp$\sigma$} \nl
\colhead{ } &                       
\colhead{} &                       
\colhead{Cepheids\tablenotemark{3}} &                      
\colhead{TRGB\tablenotemark{3}} &                      
\colhead{PNLF\tablenotemark{3}} &                      
\colhead{$V$-GCLF\tablenotemark{3}} &                      
\colhead{$B$-GCLF\tablenotemark{3}} &                      
\colhead{$I$-SBF\tablenotemark{3,4}} &                      
\colhead{F814W-SBF\tablenotemark{3}} &                     
\colhead{$K_s$-SBF\tablenotemark{3}} &                      
\colhead{$K'$-SBF\tablenotemark{3}} &
\colhead{MEAN\tablenotemark{5}}
}                        
\startdata                        
Local Group  &  & & & & & & & & &  \nl
\tableline
\hskip .1in AndV & 1 & \nd & 24.66$\pm$0.13 & \nd & \nd & \nd & \nd & \nd & \nd & \nd & 24.66$\pm$0.13 \nl
\hskip .1in AndVI & 1 & \nd & 24.61$\pm$0.12 & \nd & \nd & \nd & \nd & \nd & \nd & \nd & 24.61$\pm$0.12 \nl
\hskip .1in GR8 & 1 & $\leq$26.8: & \nd & \nd & \nd & \nd & \nd & \nd & \nd & \nd & \nd \nl
\hskip .1in IC10 & 1 & 24.10$\pm$0.19 & 24.17$\pm$0.39 & \nd & \nd & \nd & \nd & \nd & \nd & \nd & 24.11$\pm$0.17 \nl
\hskip .1in IC1613 & 1 & 24.42$\pm$0.13 & 24.26$\pm$0.16 & \nd & \nd & \nd & \nd & \nd & \nd & \nd & 24.36$\pm$0.10 \nl
\hskip .1in LeoI & 1 & \nd & 21.99$\pm$0.12 & \nd & \nd & \nd & \nd & \nd & \nd & \nd & 21.99$\pm$0.12 \nl
\hskip .1in LeoA & 1 & \nd & 24.52$\pm$0.21 & \nd & \nd & \nd & \nd & \nd & \nd & \nd & 24.52$\pm$0.21 \nl
\hskip .1in LeoB & 1 & \nd & 21.73$\pm$0.21 & \nd & \nd & \nd & \nd & \nd & \nd & \nd & 21.73$\pm$0.21 \nl
\hskip .1in LGS3 & 1 & \nd & 24.58$\pm$0.21 & \nd & \nd & \nd & \nd & \nd & \nd & \nd & 24.58$\pm$0.21 \nl
\hskip .1in LMC & 1 & 18.50$\pm$0.13 & 18.46$\pm$0.12 & 18.53$\pm$0.14 & \nd & \nd & \nd & \nd & \nd & \nd & 18.49$\pm$0.07 \nl
\hskip .1in N205 & 1 & \nd & 24.30$\pm$0.21 & 24.68$\pm$0.51 & \nd & \nd & \nd & \nd & \nd & \nd & 24.36$\pm$0.20 \nl
\hskip .1in N221 & 1 & \nd & \nd & 24.83$\pm$1.10 & \nd & \nd & 24.55$\pm$0.17 & 24.56$\pm$0.17 & \nd & 24.31:$\pm$0.14 & 24.56$\pm$0.17 \nl
\hskip .1in N224 & 1 & 24.44$\pm$0.10 & 24.45$\pm$0.19 & 24.45$\pm$0.13 & 24.34$\pm$0.30 & \nd & 24.46$\pm$0.14 & 24.84$\pm$0.15 & \nd & 24.56:$\pm$0.14 & 24.51$\pm$0.06 \nl
\hskip .1in N598 & 1 & 24.64$\pm$0.09 & 24.93$\pm$0.18 & \nd & \nd & \nd & \nd & \nd & \nd & \nd & 24.70$\pm$0.08 \nl
\hskip .1in N3109 & 1 & 25.26$\pm$0.22 & 25.60$\pm$0.09 & $\le$26.26:$\pm$0.41 & \nd & \nd & \nd & \nd & \nd & \nd & 25.56$\pm$0.08 \nl
\hskip .1in N6822 & 1 & 23.49$\pm$0.09 & 23.39$\pm$0.16 & \nd & \nd & \nd & \nd & \nd & \nd & \nd & 23.47$\pm$0.08 \nl
\hskip .1in SextansA & 1 & 25.85$\pm$0.15 & 25.70$\pm$0.11 & \nd & \nd & \nd & \nd & \nd & \nd & \nd & 25.76$\pm$0.09 \nl
\hskip .1in SextansB & 1 & 25.69$\pm$0.25 & 25.60$\pm$0.12 & \nd & \nd & \nd & \nd & \nd & \nd & \nd & 25.62$\pm$0.11 \nl
\hskip .1in SMC & 1 & 18.99$\pm$0.05 & \nd & 19.29:$\pm$0.31 & \nd & \nd & \nd & \nd & \nd & \nd & 18.99$\pm$0.05 \nl
\hskip .1in WLM & 1 & \nd & 24.84$\pm$0.12 & \nd & \nd & \nd & \nd & \nd & \nd & \nd & 24.84$\pm$0.12 \nl
\tableline
\tableline
N1023 Group &  1  &   29.84$\pm$0.08  &  \nd  &  30.08$\pm$0.08  &  \nd  &  \nd  &  30.08$\pm$0.11  &  30.44$\pm$0.14  &  \nd  &  \nd  &  30.03$\pm$0.05 \nl
\multicolumn{2}{l}{Depth 0.32 mag} & & & & & & & & \nl
\tableline
\hskip .1in N891 & 1 & \nd & \nd & 30.10$\pm$0.13 & \nd & \nd & 29.67$\pm$0.18 & \nd & \nd & \nd & 29.95$\pm$0.10 \nl
\hskip .1in N925 & 1 & 29.84$\pm$0.08 & \nd & \nd & \nd & \nd & \nd & \nd & \nd & \nd & 29.84$\pm$0.08 \nl
\hskip .1in N1023 & 1 & \nd & \nd & 30.07$\pm$0.11 & \nd & \nd & 30.35$\pm$0.19 & 30.44$\pm$0.14 & \nd & \nd & 30.23$\pm$0.08 \nl
\tableline
\tableline
Fornax Cluster & 1 & 31.60$\pm$0.04 &  \nd & 31.20$\pm$0.07 & 31.38$\pm$0.15 & 31.76$\pm$0.31 & 31.59$\pm$0.04 &  \nd &  \nd & 31.51:$\pm$0.08 & 31.52$\pm$0.03\nl
\multicolumn{2}{l}{Depth 0.28 mag} & & & & & & & & \nl
\tableline
\hskip .1in N1316 & 1 & \nd & \nd & 31.15$\pm$0.11 & \nd & \nd & 31.71$\pm$0.19 & \nd & \nd & \nd & 31.28$\pm$0.09 \nl
\hskip .1in N1326A & 1 & 31.43$\pm$0.07 & \nd & \nd & \nd & \nd & \nd & \nd & \nd & \nd & 31.43$\pm$0.07 \nl
\hskip .1in N1339 & 1 & \nd & \nd & \nd & \nd & \nd & 31.66$\pm$0.37 & \nd & \nd & 31.73:$\pm$0.18 & 31.66$\pm$0.37 \nl
\hskip .1in N1344 & 1 & \nd & \nd & \nd & 31.34$\pm$0.35 & \nd & 31.53$\pm$0.31 & \nd & \nd & 31.35:$\pm$0.16 & 31.53$\pm$0.31 \nl
\hskip .1in N1365 & 1 & 31.39$\pm$0.10 & \nd & \nd & \nd & \nd & \nd & \nd & \nd & \nd & 31.39$\pm$0.10 \nl
\hskip .1in N1379 & 1 & \nd & \nd & \nd & \nd & 31.92$\pm$0.58 & 31.56$\pm$0.17 & \nd & \nd & 31.21:$\pm$0.17 & 31.56$\pm$0.17 \nl
\hskip .1in N1380 & 1 & \nd & \nd & \nd & 31.28$\pm$0.27 & \nd & 31.37$\pm$0.20 & \nd & \nd & \nd & 31.37$\pm$0.20 \nl
\hskip .1in N1399 & 1 & \nd & \nd & 31.22$\pm$0.11 & 31.31$\pm$0.27 & 31.56$\pm$0.51 & 31.56$\pm$0.17 & \nd & \nd & 31.75:$\pm$0.19 & 31.31$\pm$0.09 \nl
\hskip .1in N1404 & 1 & \nd & \nd & 31.22$\pm$0.12 & 31.59$\pm$0.29 & 31.85$\pm$0.54 & 31.64$\pm$0.20 & \nd & \nd & 31.59:$\pm$0.18 & 31.33$\pm$0.10 \nl
\hskip .1in N1425 & 1 & 31.81$\pm$0.06 & \nd & \nd & \nd & \nd & \nd & \nd & \nd & \nd & 31.81$\pm$0.06 \nl
\tableline
\tableline
Eridanus Cluster & 1 & \nd & \nd & \nd &  \nd &  \nd & 31.99$\pm$0.11 &  \nd &  \nd &    31.57:$\pm$0.13 & 31.99$\pm$0.11\nl
                 & 1,2 & \nd & \nd & \nd &  \nd &  \nd & 32.01$\pm$0.10 &  \nd &  \nd &    31.69:$\pm$0.11 & 32.01$\pm$0.10 \nl
\multicolumn{2}{l}{Depth 0.35 mag} & & & & & & & & \nl
\tableline
\hskip .1in N1395 & 1 & \nd & \nd & \nd & \nd & \nd & 31.97$\pm$0.18 & \nd & \nd & 31.20:$\pm$0.20 & 31.97$\pm$0.18 \nl
\hskip .1in N1400 & 2 & \nd & \nd & \nd & \nd & \nd & 32.19$\pm$0.33 & \nd & \nd & 32.22:$\pm$0.27 & 32.19$\pm$0.33 \nl
\hskip .1in N1407 & 1 & \nd & \nd & \nd & \nd & \nd & 32.30$\pm$0.26 & \nd & \nd & 31.95:$\pm$0.19 & 32.30$\pm$0.26 \nl
\hskip .1in N1426 & 1 & \nd & \nd & \nd & \nd & \nd & 31.89$\pm$0.16 & \nd & \nd & 31.44:$\pm$0.32 & 31.89$\pm$0.16 \nl
\tableline
\tableline
M81 Group & 1 & 27.80$\pm$0.08 & 27.88$\pm$0.08 & 27.87$\pm$0.12 &  \nd &  \nd & 28.07$\pm$0.15 &  \nd &  \nd &  \nd & 27.87$\pm$0.05\nl
\multicolumn{2}{l}{Depth 0.15 mag} & & & & & & & & \nl
\tableline
\hskip .1in N2366 & 2 & $\leq$27.7: & \nd & \nd & \nd & \nd & \nd & \nd & \nd & \nd & \nd \nl
\hskip .1in N2403 & 2 & $\leq$27.5: & \nd & \nd & \nd & \nd & \nd & \nd & \nd & \nd & \nd \nl
\hskip .1in N3031 & 1 & 27.80$\pm$0.08 & \nd & 27.88$\pm$0.12 & \nd & \nd & 28.02$\pm$0.28 & \nd & \nd & \nd & 27.84$\pm$0.06 \nl
\hskip .1in BK5N & 1 & \nd & 27.88$\pm$0.13 & \nd & \nd & \nd & \nd & \nd & \nd & \nd & 27.88$\pm$0.13 \nl
\hskip .1in F8DI & 1 & \nd & 27.86$\pm$0.11 & \nd & \nd & \nd & \nd & \nd & \nd & \nd & 27.86$\pm$0.11 \nl
\hskip .1in U4305 & 1  & $\leq$27.42:  & \nd & \nd & \nd & \nd & \nd & \nd & \nd & \nd & \nd \nl 
\tableline
\tableline
N3184 Group & 1 & 30.79$\pm$0.05 &  \nd &  \nd &  \nd &  \nd &  \nd &  \nd &  \nd &  \nd & 30.80$\pm$0.06\nl
\multicolumn{2}{l}{Depth 0.35 mag} & & & & & & & & \nl
\tableline
\hskip .1in N3198 & 1 & 30.80$\pm$0.06 & \nd & \nd & \nd & \nd & \nd & \nd & \nd & \nd & 30.80$\pm$0.06 \nl
\hskip .1in N3319 & 1 & 30.78$\pm$0.10 & \nd & \nd & \nd & \nd & \nd & \nd & \nd & \nd & 30.78$\pm$0.10 \nl
\tableline
\tableline
Leo I Group & 1,2 & 30.08$\pm$0.06 & 30.33$\pm$0.11 & 30.01$\pm$0.07 &  \nd & 29.62$\pm$0.56 & 30.30$\pm$0.06 & 30.26$\pm$0.07 & 30.48:$\pm$0.14 &  \nd & 30.18$\pm$0.03\nl
\multicolumn{2}{l}{Depth 0.15 mag} & & & & & & & & \nl
\tableline
\hskip .1in N3351 & 1 & 30.01$\pm$0.08 & \nd & \nd & \nd & \nd & \nd & \nd & \nd & \nd & 30.01$\pm$0.08 \nl
\hskip .1in N3368 & 1 & 30.20$\pm$0.10 & \nd & 29.91$\pm$0.12 & \nd & \nd & 30.14$\pm$0.23 & \nd & \nd & \nd & 30.09$\pm$0.07 \nl
\hskip .1in N3377 & 1 & \nd & \nd & 30.10$\pm$0.19 & \nd & \nd & 30.31$\pm$0.12 & 30.12$\pm$0.12 & \nd & \nd & 30.19$\pm$0.08 \nl
\hskip .1in N3379 & 1 & \nd & 30.33$\pm$0.11 & 30.01$\pm$0.15 & \nd & 29.62$\pm$0.56 & 30.18$\pm$0.14 & 30.31$\pm$0.12 & 30.48:$\pm$0.14 & \nd & 30.23$\pm$0.06 \nl
\hskip .1in N3384 & 1 & \nd & \nd & 30.09$\pm$0.15 & \nd & \nd & 30.38$\pm$0.16 & 30.34$\pm$0.11 & \nd & \nd & 30.28$\pm$0.08 \nl
\hskip .1in N3627 & 2 & 30.06$\pm$0.17 & \nd & \nd & \nd & \nd & \nd & \nd & \nd & \nd & 30.06$\pm$0.17 \nl
\tableline
\tableline
Coma I Cloud & 2 & 31.41$\pm$0.10 &  \nd & 30.12$\pm$0.18 & 30.74$\pm$0.28 &  \nd & 30.79$\pm$0.09 &  \nd &  \nd &  \nd & 30.96$\pm$0.06\nl
\multicolumn{2}{l}{Depth 0.52 mag} & & & & & & & & \nl
\tableline
\hskip .1in N4278 & 2 & \nd & \nd & 30.12$\pm$0.18 & 30.74$\pm$0.28 & \nd & 31.09$\pm$0.21 & \nd & \nd & \nd & 30.51$\pm$0.14 \nl
\hskip .1in N4414 & 2 & 31.41$\pm$0.10 & \nd & \nd & \nd & \nd & \nd & \nd & \nd & \nd & 31.41$\pm$0.10 \nl
\tableline
\tableline
Coma II Cloud & 2 & 30.57$\pm$0.08 &  \nd & 30.51$\pm$0.08 & 30.44$\pm$0.21 &  \nd & 31.18$\pm$0.10 &  \nd &  \nd &  \nd & 30.68$\pm$0.05\nl
\multicolumn{2}{l}{Depth 0.52 mag} & & & & & & & & \nl
\tableline
\hskip .1in N4494 & 2 & \nd & \nd & 30.61$\pm$0.09 & 30.64$\pm$0.28 & \nd & 31.22$\pm$0.13 & \nd & \nd & \nd & 30.81$\pm$0.08 \nl
\hskip .1in N4565 & 2 & \nd & \nd & 30.20$\pm$0.16 & 30.18$\pm$0.32 & \nd & 31.26$\pm$0.19 & \nd & \nd & \nd & 30.63$\pm$0.12 \nl
\hskip .1in N4725 & 2 & 30.57$\pm$0.08 & \nd & \nd & \nd & \nd & 30.67$\pm$0.35 & \nd & \nd & \nd & 30.57$\pm$0.08 \nl
\tableline
\tableline
Virgo Cluster & 1 & 31.04$\pm$0.06 & 30.91$\pm$0.10 & 30.90$\pm$0.07 & 31.29$\pm$0.23 &  \nd & 31.19$\pm$0.04 & 31.09$\pm$0.06 & 30.87:$\pm$0.09 & 31.18:$\pm$0.10 & 31.07$\pm$0.03\nl
\multicolumn{2}{l}{M87 subcluster} & & & & & & & & \nl
\multicolumn{2}{l}{Depth 0.21 mag} & & & & & & & & \nl
\tableline
\hskip .1in IC3388 & 1 & \nd & 30.84$\pm$0.10 & \nd & \nd & \nd & \nd & \nd & \nd & \nd & 30.84$\pm$0.10 \nl
\hskip .1in IG Stars & 1 & \nd & 31.44$\pm$0.31 & \nd & \nd & \nd & \nd & \nd & \nd & \nd & 31.44$\pm$0.31 \nl
\hskip .1in N4321 & 1 & 31.04$\pm$0.09 & \nd & \nd & \nd & \nd & \nd & \nd & \nd & \nd & 31.04$\pm$0.09 \nl
\hskip .1in N4374 & 1 & \nd & \nd & 30.99$\pm$0.15 & \nd & \nd & 31.40$\pm$0.14 & \nd & 31.42:$\pm$0.25 & \nd & 31.20$\pm$0.10 \nl
\hskip .1in N4382 & 1 & \nd & \nd & 30.83$\pm$0.14 & \nd & \nd & 31.39$\pm$0.17 & \nd & \nd & \nd & 31.07$\pm$0.11 \nl
\hskip .1in N4406 & 1 & \nd & \nd & 31.03$\pm$0.13 & \nd & \nd & 31.33$\pm$0.12 & 31.27$\pm$0.12 & \nd & 31.22:$\pm$0.15 & 31.22$\pm$0.07 \nl
\hskip .1in N4458 & 1 & \nd & \nd & \nd & \nd & \nd & 31.17$\pm$0.24 & 30.92$\pm$0.13 & \nd & \nd & 30.98$\pm$0.11 \nl
\hskip .1in N4473 & 1 & \nd & \nd & \nd & \nd & \nd & 31.04$\pm$0.16 & 31.11$\pm$0.13 & \nd & \nd & 31.08$\pm$0.10 \nl
\hskip .1in N4478 & 1 & \nd & \nd & \nd & 31.34$\pm$0.46 & \nd & 31.35$\pm$0.29 & \nd & 31.24:$\pm$0.17 & \nd & 31.35$\pm$0.29 \nl
\hskip .1in N4486 & 1 & \nd & \nd & 30.82$\pm$0.12 & 31.27$\pm$0.26 & \nd & 31.22$\pm$0.16 & \nd & \nd & \nd & 30.97$\pm$0.10 \nl
\hskip .1in N4489 & 1 & \nd & \nd & \nd & \nd & \nd & 31.31$\pm$0.18 & \nd & 29.56:$\pm$0.25 & 30.75:$\pm$0.37 & 31.31$\pm$0.18 \nl
\hskip .1in N4548 & 1 & 31.04$\pm$0.08 & \nd & \nd & \nd & \nd & 31.48$\pm$0.55 & \nd & \nd & \nd & 31.05$\pm$0.08 \nl
\hskip .1in N4552 & 1 & \nd & \nd & \nd & \nd & \nd & 30.99$\pm$0.17 & 31.02$\pm$0.14 & 30.85:$\pm$0.13 & 31.21:$\pm$0.15 & 31.01$\pm$0.11 \nl
\hskip .1in N4571 & 1 & $\leq$30.9: & \nd & \nd & \nd & \nd & \nd & \nd & \nd & \nd & \nd \nl
\tableline
\tableline
Virgo Cluster & 1 & 31.02$\pm$0.04 &  \nd & 30.79$\pm$0.15 & 31.30$\pm$0.27 & 31.71$\pm$0.55 & 31.07$\pm$0.08 & 31.24$\pm$0.13 & 31.01:$\pm$0.12 & 31.18:$\pm$0.12 & 31.03$\pm$0.03\nl
\multicolumn{2}{l}{NGC 4472 subcluster} & & & & & & & & \nl
\multicolumn{2}{l}{Depth 0.21 mag} & & & & & & & & \nl
\tableline
\hskip .1in N4472 & 1 & \nd & \nd & 30.79$\pm$0.15 & 31.30$\pm$0.27 & 31.71$\pm$0.55 & 31.11$\pm$0.13 & 31.24$\pm$0.13 & 31.25:$\pm$0.17 & 31.07:$\pm$0.15 & 31.07$\pm$0.08 \nl
\hskip .1in N4496A & 1 & 31.02$\pm$0.07 & \nd & \nd & \nd & \nd & \nd & \nd & \nd & \nd & 31.02$\pm$0.07 \nl
\hskip .1in N4526 & 1 & \nd & \nd & \nd & \nd & \nd & 31.19$\pm$0.22 & \nd & \nd & \nd & 31.19$\pm$0.22 \nl
\hskip .1in N4535 & 1 & 31.10$\pm$0.07 & \nd & \nd & \nd & \nd & \nd & \nd & \nd & \nd & 31.10$\pm$0.07 \nl
\hskip .1in N4536 & 1 & 30.95$\pm$0.07 & \nd & \nd & \nd & \nd & \nd & \nd & \nd & \nd & 30.95$\pm$0.07 \nl
\hskip .1in N4636 & 1 & \nd & \nd & \nd & \nd & \nd & 30.88$\pm$0.16 & \nd & 30.79:$\pm$0.16 & 31.35:$\pm$0.18 & 30.88$\pm$0.16 \nl
\tableline
\tableline
Virgo Cluster & 1 & 31.80$\pm$0.09 &  \nd & 30.82$\pm$0.15 &  \nd & 31.38$\pm$0.53& 31.22$\pm$0.07 & 30.98$\pm$0.08 & 31.41:$\pm$0.19 & 31.31:$\pm$0.32 & 31.25$\pm$0.04\nl
\multicolumn{2}{l}{NGC 4649 subcluster} & & & & & & & & \nl
\multicolumn{2}{l}{Depth 0.21 mag} & & & & & & & & \nl
\tableline
\hskip .1in N4578 & 1 & \nd & \nd & \nd & \nd & \nd & 31.40$\pm$0.15 & \nd & \nd & 31.31:$\pm$0.32 & 31.40$\pm$0.15 \nl
\hskip .1in N4621 & 1 & \nd & \nd & \nd & \nd & \nd & 31.37$\pm$0.22 & 30.85$\pm$0.13 & 31.41:$\pm$0.19 & \nd & 30.99$\pm$0.11 \nl
\hskip .1in N4639 & 1 & 31.80$\pm$0.09 & \nd & \nd & \nd & \nd & \nd & \nd & \nd & \nd & 31.80$\pm$0.09 \nl
\hskip .1in N4649 & 1 & \nd & \nd & 30.82$\pm$0.15 & \nd & 31.38$\pm$0.53 & 31.18$\pm$0.17 & 31.23$\pm$0.13 & \nd & \nd & 31.09$\pm$0.08 \nl
\hskip .1in N4660 & 1 & \nd & \nd & \nd & \nd & \nd & 30.61$\pm$0.21 & 30.76$\pm$0.18 & \nd & \nd & 30.70$\pm$0.14 \nl
\tableline
\tableline
Cen A Group & 1 & 27.61$\pm$0.11 & 27.93$\pm$0.13 & 27.70$\pm$0.11 &  \nd &  \nd & 28.12$\pm$0.13 &  \nd &  \nd &  \nd & 27.82$\pm$0.06\nl
\multicolumn{2}{l}{Depth 0.67 mag} & & & & & & & & \nl
\tableline
\hskip .1in N5102 & 1 & \nd & \nd & 27.54$\pm$0.22 & \nd & \nd & 28.07$\pm$0.17 & \nd & \nd & \nd & 27.89$\pm$0.13 \nl
\hskip .1in N5128 & 1 & \nd & 27.93$\pm$0.13 & 27.76$\pm$0.12 & \nd & \nd & 28.19$\pm$0.19 & \nd & \nd & \nd & 27.90$\pm$0.08 \nl
\hskip .1in N5253 & 1 & 27.61$\pm$0.11 & \nd & 28.13:$\pm$0.41 & \nd & \nd & \nd & \nd & \nd & \nd & 27.61$\pm$0.11 \nl
\tableline
\tableline
M101 Group & 1 & 29.34$\pm$0.10 &  \nd & 29.42$\pm$0.13 &  \nd &  \nd &  \nd &  \nd &  \nd &  \nd & 29.37$\pm$0.08\nl
& 1+3 & 29.34$\pm$0.10  \nd & \nd & 29.47$\pm$0.09 &  \nd &  \nd & 30.67$\pm$0.13 &  \nd &  \nd &  \nd & 29.69$\pm$0.06\nl
\multicolumn{2}{l}{Depth 0.35 mag} & & & & & & & & \nl
\tableline
\hskip .1in N5195 & 3 & \nd & \nd & 29.52$\pm$0.12 & \nd & \nd & 29.47$\pm$0.28 & \nd & \nd & \nd & 29.52$\pm$0.11 \nl
\hskip .1in N5457 & 1 & 29.34$\pm$0.10 & \nd & 29.42$\pm$0.13 & \nd & \nd & \nd & \nd & \nd & \nd & 29.37$\pm$0.08 \nl
\tableline
\tableline
N7331 Group & 1 & 30.89$\pm$0.10 &  \nd &  \nd &  \nd &  \nd & 30.67$\pm$0.21 &  \nd &  \nd &  \nd & 30.85$\pm$0.09\nl
            & 1+2 & 30.89$\pm$0.10  &  \nd &  \nd &  \nd & \nd & 30.67$\pm$0.16 & 30.76$\pm$0.15 &  \nd &  \nd & 30.81$\pm$0.07\nl
\multicolumn{2}{l}{Depth 0.35 mag} & & & & & & & & \nl
\tableline
\hskip .1in N7331 & 1 & 30.89$\pm$0.10 & \nd & \nd & \nd & \nd & 30.67$\pm$0.21 & \nd & \nd & \nd & 30.85$\pm$0.09 \nl
\hskip .1in N7457 & 2 & \nd & \nd & \nd & \nd & \nd & 30.66$\pm$0.23 & 30.76$\pm$0.15 & \nd & \nd & 30.73$\pm$0.13 \nl
\tableline
\tableline
Other Groups  &  & & & & & & & & &  \nl
\tableline
\hskip .1in IC4182 & 1 & 28.36$\pm$0.08 & \nd & \nd & \nd & \nd & \nd & \nd & \nd & \nd & 28.36$\pm$0.08 \nl
\hskip .1in N300 & 1 & 26.66$\pm$0.10 & \nd & 26.93$\pm$0.41 & \nd & \nd & \nd & \nd & \nd & \nd & 26.68$\pm$0.10 \nl
\hskip .1in N2090 & 1 & 30.45$\pm$0.08 & \nd & \nd & \nd & \nd & \nd & \nd & \nd & \nd & 30.45$\pm$0.08 \nl
\hskip .1in N2541 & 1 & 30.47$\pm$0.08 & \nd & \nd & \nd & \nd & \nd & \nd & \nd & \nd & 30.47$\pm$0.08 \nl
\hskip .1in N3115 & 1 & \nd & 30.18$\pm$0.30 & 30.12$\pm$0.19 & 29.90$\pm$0.26 & \nd & 29.99$\pm$0.13 & \nd & \nd & \nd & 30.05$\pm$0.10 \nl
\hskip .1in N3115DW1 & 1 & \nd & \nd & \nd & 30.53$\pm$0.84 & \nd & \nd & \nd & \nd & \nd & \nd \nl
\hskip .1in N3621 & 1 & 29.13$\pm$0.11 & \nd & \nd & \nd & \nd & \nd & \nd & \nd & \nd & 29.13$\pm$0.11 \nl
\hskip .1in N4365 & 1 & \nd & \nd & \nd & 31.80$\pm$0.39 & 32.31$\pm$0.68 & 31.61$\pm$0.19 & \nd & 31.14:$\pm$0.16 & 31.86:$\pm$0.15 & 31.61$\pm$0.19 \nl
\hskip .1in N4373 & 1 & \nd & \nd & \nd & \nd & \nd & 33.23$\pm$0.22 & 32.76$\pm$0.23 & \nd & \nd & 33.01$\pm$0.16 \nl
\hskip .1in N4594 & 1 & \nd & \nd & 29.75$\pm$0.11 & \nd & \nd & 30.01$\pm$0.20 & 29.99$\pm$0.17 & \nd & \nd & 29.85$\pm$0.08 \nl
\hskip .1in N4603 & 1 & \nd & \nd & \nd & \nd & \nd & \nd & \nd & \nd & \nd & \nd \nl
\hskip .1in N5170 & 1 & \nd & \nd & \nd & 31.50$\pm$0.34 & \nd & \nd & \nd & \nd & \nd & \nd \nl
\hskip .1in N5481 & 1 & \nd & \nd & \nd & 31.84$\pm$0.56 & \nd & \nd & \nd & \nd & \nd & \nd \nl
\hskip .1in N5846 & 1 & \nd & \nd & \nd & 32.61$\pm$0.28 & \nd & 32.05$\pm$0.22 & \nd & \nd & \nd & 32.05$\pm$0.22 \nl
\enddata 
\tablenotetext{1}{References for the original data from which the distances given in this table were derived can be found in Ferrarese et al. (1999).}
\tablenotetext{2}{Galaxies which are certain, very likely, or just probable
members of their assigned group or cluster are designated as class 1,
2, and 3 respectively.}
\tablenotetext{3}{The errors quoted on the distance moduli are random errors only, 
with the exception of the ground-based Cepheid distances which include both random and systematic errors. Systematic errors amount
to 0.13 mag for TRGB and PNLF distance moduli, and 0.16 mag for GCLF, $I$-, F814W-, $K_s$-, and $K'$-SBF, and
all Cepheid distance moduli observed with the $HST$}
\tablenotetext{4}{All $I$-SBF distances quoted from this table must reference Tonry et al. (1999) and Ajhar et al. (1999) as the
source of the SBF measurement, combined with the calibration presented in this paper.}
\tablenotetext{5}{Weighted mean of all of the available Cepheid, PNLF, TRGB, $I$- and F814W-SBF. GCLF and near infrared SBF measurements are not included for the reasons discussed in \S 5 and Appendix B.}
\end{deluxetable}